\documentclass[11pt,a4paper]{article}
\usepackage{jheppub}

\usepackage{amssymb,amsmath,amstext,amsfonts,empheq}
\usepackage{bbold}
\usepackage{bm}
\usepackage{graphicx}
\usepackage{xcolor}
\usepackage{overpic}
\usepackage{subcaption}

\usepackage{cancel}
\usepackage{soul}
\usepackage{bm}

\usepackage[normalem]{ulem}
\usepackage{hyperref}
\definecolor{rossos}{cmyk}{0,1,1,0.55}
\definecolor{bluc}{cmyk}{1,1,0,0.1}
\definecolor{blu}{cmyk}{1,1,0,0.3}
\hypersetup{colorlinks, bookmarksopen, bookmarksnumbered,
citecolor=blu, linkcolor=bluc, urlcolor=rossos}

\def\ii{{\rm i}} 

\newcommand{\bx}{{\bf x}}
\newcommand{\bn}{{\bf n}}
\newcommand*\boldell{\ensuremath{\boldsymbol\ell}}

\newcommand{\nn}{\nonumber}
\newcommand{\beq}{\begin{equation}}
\newcommand{\eeq}{\end{equation}}

\newcommand{\bea}{\begin{eqnarray}}
\newcommand{\eea}{\end{eqnarray}}
\newcommand{\be}{\begin{equation}}
\newcommand{\ee}{\end{equation}}

\newcommand\Fresponse{{\cal T}}
\newcommand{\dd}[0]{\mathrm{d}}

\title{\centering
  Probing the Kinematic Dipole with LISA: \\ an analytical treatment}
\author[a]{Jacopo Fumagalli,}
\author[b,c]{Giulia Cusin,}
\author[b]{Cyril Pitrou}
\author[d,e]{and Gianmassimo Tasinato}

\affiliation[a]{Departement de F\'isica Qu\`antica i Astrofisica and Institut de Ci\`encies del Cosmos (ICC), Universitat de Barcelona, Mart\'i i Franqu\`es 1, 08028 Barcelona, Spain}
\affiliation[b]{Institut d'Astrophysique de Paris, UMR-7095 du CNRS et de Sorbonne Universit\'e, Paris, France}
\affiliation[c]{Département de Physique Théorique and Center for Astroparticle Physics, Université de Genève, Quai E. Ansermet 24, CH-1211 Genève 4, Switzerland}
\affiliation[d]{Physics Department, Swansea University, Singleton Campus, 12 SA2 8PP, U.K}
\affiliation[e]{Dipartimento di Fisica e Astronomia, Universit\`a di Bologna,\\ INFN, Sezione di Bologna, viale B. Pichat 6/2, 40127 Bologna, Italy}

\emailAdd{jfumagalli@fqa.ub.edu}\emailAdd{cusin@iap.fr}
\emailAdd{pitrou@iap.fr}
\emailAdd{g.tasinato2208.at.gmail.com}
\abstract{
The motion of the Solar System with respect to the cosmic rest frame induces a kinematic dipole in the stochastic gravitational-wave background (GWB). Detecting this signal with space-based interferometers would provide an independent measurement of our peculiar velocity and a GW probe of cosmic anisotropies. We present a fully analytic derivation of the response of the \emph{Laser Interferometer Space Antenna} (LISA) to a kinematic dipole, and construct an optimal estimator for its detection. We show that the dipolar response is governed by a single frequency-dependent function fixed by symmetry, and we compute its behaviour across the LISA band. Using Fisher forecasts, we find that for a scale-invariant background detectability requires $h^2\Omega_{\rm GW} \gtrsim 5\times 10^{-8}$ for \emph{fiducial} LISA, and $h^2\Omega_{\rm GW} \gtrsim 5\times 10^{-10}$ for a detector with characteristic instrumental-noise amplitudes improved by an order of magnitude. Prospects are more favorable for signals with richer frequency profile. We also explore the potential of the kinematic dipole to break degeneracies, particularly in the presence of strong galactic foregrounds or noise features that closely mimic the primordial signal.}

\begin{document}
\hfill{\flushright {}}
\maketitle
\newpage
\section{Introduction}

The Solar System moves with a velocity $\beta = v/c \simeq 1.23 \times 10^{-3}$ with respect 
to the rest frame of the cosmic microwave background (CMB),  towards the direction
$(\ell, b) = (264^\circ, 48^\circ)$ in Galactic coordinates.
 This motion is responsible 
for the large dipolar anisotropy observed in CMB temperature maps \cite{Fixsen:1993rd,Fixsen:1996nj,Planck:2018vyg,Planck:2013kqc}. 
Following the suggestion of \cite{Ellis:1984uka}, kinematic anisotropies are increasingly being probed through astrophysical measurements, in some cases with results in tension with the value inferred from the CMB. 
See \cite{Blake:2002gx,Singal_2011,Gibelyou:2012ri,Rubart:2013tx} for seminal works on the subject; recent studies include radio-source analyses \cite{Secrest:2020has,Dam:2022wwh,Secrest:2022uvx,Bohme:2025nvu,daSilveiraFerreira:2024ddn}, supernovae Ia \cite{Horstmann:2021jjg}, and optical galaxies \cite{Tiwari:2024dsj}.
While supernovae Ia and optical-galaxy measurements show no comparable dipole anomaly, radio-source measurements indicate a discrepancy at more than five sigma. Recent reviews are provided in \cite{Aluri:2022hzs,Secrest:2025wyu}.

This discrepancy could have several possible origins. It may arise from systematic effects in the quasar and radio-source data sets; see, e.g., \cite{vonHausegger:2025iuo}. Another proposed explanation is imperfect theoretical modelling of the expected number-count dipole, in particular the neglect of evolution effects~\citep{Dalang:2021ruy,Guandalin:2022tyl}, although these were later found not to affect the dipole when source properties are estimated near the flux limit~\cite{vonHausegger:2024jan}, unless additional selection effects are included~\cite{vonHausegger:2024fcu}. A more radical possibility is a violation of cosmic isotropy, in which case the large dipole observed in quasars and radio sources would not be due solely to the observer’s velocity, but would also contain an intrinsic contribution generated by a large local anisotropy, in tension with $\Lambda$CDM predictions; for early constraints on this scenario using CMB data, see \cite{Ferreira:2020aqa,Ferreira:2021omv}.

One way to test these scenarios is to use other datasets to measure the dipole at different redshifts, and gravitational wave observables may provide an interesting and independent probe of these effects, see \cite{Cusin:2024git} for a review of kinematic effects on GW observables. In \cite{Mastrogiovanni:2022nya, Grimm:2023tfl}, it has been suggested that an independent measurement of our peculiar motion can be achieved by looking at the distribution of GW events in the context of next-generation gravitational wave detectors. Another interesting possibility is to look for anisotropies in the stochastic gravitational wave background (GWB) \cite{LISACosmologyWorkingGroup:2022kbp,Cusin:2022cbb}. It is therefore important to forecast the capabilities of forthcoming and future experiments to detect such kinematic anisotropies \cite{Domcke:2019zls,ValbusaDallArmi:2022htu,Chung:2022xhv,Chowdhury:2022pnv,Mastrogiovanni:2022nya,Mentasti:2023gmg,Tasinato:2023zcg,Agarwal:2023lzz,NANOGrav:2023tcn,Cruz:2024diu,Cruz:2024esk,Cusin:2024git,Cruz:2024svc,Ebersold:2024hgp,Li:2024lvt,Depta:2024ykq,Mentasti:2025ywl,LIGOScientific:2025bkz,Blumke:2025nrq,AnilKumar:2026hrn}, and to assess their ability to distinguish them from intrinsic anisotropies in the GWB—see, e.g., \cite{Contaldi:2016koz,Bartolo:2019oiq,Bartolo:2019yeu,Adshead:2020bji} for the theoretical characterization of the latter.
Observationally, since the strain associated with a stochastic background is expected to resemble white noise and is typically below the instrumental noise threshold, its detection requires correlating the output of at least two independent detectors, under the assumption that their noise is uncorrelated. A key difference between ground-based detectors and space-based missions such as \emph{Laser Interferometer Space Antenna} (LISA) \cite{LISA:2024hlh} lies in the limited knowledge of the noise. In LISA, noise is correlated among the three nested interferometers, and, because the mission operates in a strong-signal regime, it is not possible to estimate the noise using data “far from the event” \cite{PhysRevD.109.042001}. One possible approach is to exploit the so-called null channel, although this only provides an upper bound on the noise in the signal channels \cite{Muratore:2022nbh}. Additionally, irreducible galactic foregrounds represent a further challenge for space-based missions, as they can obscure primordial signal components.

In this context, we investigate the capability of LISA to detect and characterize the kinematic dipole of the GWB. Beyond providing an independent measurement of our peculiar velocity, the kinematic dipole offers a distinctive handle to break degeneracies: since neither instrumental noise nor galactic foregrounds are expected to exhibit kinematic anisotropies\footnote{Galactic foregrounds do present intrinsic anisotropies, which are typically modeled and incorporated into analysis templates. We neglect this effect here, since we do not expect the intrinsic dipole to be aligned with the kinematic one, and therefore the leading-order contamination is negligible.}, its inclusion in the analysis can help disentangle primordial or extragalactic signals from contaminants. We place particular emphasis on developing a physically transparent analytic understanding of the LISA response to kinematic effects. The response of LISA to such effects presents subtle features due to the planar geometry of the detector and its time-dependent orientation. We study in detail the dependence of the response on the gravitational-wave frequency and on the annual motion of the detector \cite{Cornish:2003tz}, proposing a fully analytic approach. By developing dedicated Fisher forecasts, we assess the prospects for detecting the kinematic dipole for large GWB amplitudes motivated by early-Universe scenarios. 
For representative examples of signals at levels considered in the literature -- see the analyses 
of the LISA Cosmology Working Group \cite{Caprini:2015zlo,Bartolo:2016ami,Auclair:2019wcv,LISACosmologyWorkingGroup:2022jok,LISACosmologyWorkingGroup:2024hsc,LISACosmologyWorkingGroup:2025vdz} for 
theoretical motivations and references -- 
we find that detectability can be achieved for $\Omega_* \gtrsim 5\times 10^{-8}$ with
\emph{fiducial} LISA for a scale-invariant spectrum. These findings are consistent with the results of \cite{Heisenberg:2024var,LISACosmologyWorkingGroup:2022kbp}. Concerning the dipole detectability, we go beyond previous analysis showing that the reconstruction of our peculiar velocity can be significantly more promising for backgrounds with richer spectral shapes. We also investigate, using a similar analytical approach, the role of the kinematic quadrupole. We find that including it in the analysis does not improve the constraints on the velocity, in agreement with the results of the simulations presented \cite{Heisenberg:2024var}. 
We stress the advantage of having an analytic treatment: the underlying physics is fully transparent, and forecasts
can be derived very efficiently, without the need to run a complete numerical pipeline.
Besides being an interesting target per se, the detection of a kinematic dipole carries important physical implications. In particular, as anticipated above, it provides a complementary observable for separating cosmological or extragalactic signal profiles from astrophysical and instrumental contaminants. We show, for the first time, that incorporating kinematic information leads to improved constraints on the parameters of a cosmological stochastic background component, particularly in the presence of very large foregrounds or noise features that are highly degenerate with the signal. Finally, we provide analytic expressions for the Fisher matrix elements, which can be readily adapted to a wide range of benchmark models in future studies.

Our analytic treatment of both the detectability of the kinematic dipole and its role as a degeneracy breaker offers conceptual clarity and is intended for astrophysicists and cosmologists seeking a first-principles understanding of the problem. These results are especially relevant in view of future post-LISA missions, which are expected to face significant foreground contamination and will be the subject of further investigation in upcoming work.

The paper is structured as follows. In Section~\ref{sec_lisarfd}, we develop an analytic understanding of the response of the LISA instrument to a kinematic dipole in the stochastic gravitational-wave background. In Section~\ref{estimator}, we construct an estimator for the peculiar velocity and show that its reconstruction is only possible due to the motion of LISA. In Section~\ref{deg_breaker}, we introduce the idea that the dipole can help reconstructing signal parameters in situations where the foreground is degenerate with the signal or where noise features mimic a signal. After presenting a heuristic argument, we perform an explicit analysis through several case studies, considering both the current LISA configuration and a future instrument operating in the same frequency band with improved sensitivity.

\section{LISA response  to kinematic anisotropies}
\label{sec_lisarfd}

In this introductory section we develop an analytic understanding of the response 
of the LISA instrument to a kinematic dipole in the stochastic gravitational wave 
background. Our goal is to identify the physical effects that allow the detector 
to measure such an anisotropy, and to clarify how the detector geometry and motion 
shape the observable signal.

To do so, we extend to the case of a dipolar anisotropy the clear and pedagogical 
discussion presented in \cite{Smith:2019wny}, which was originally developed for 
an isotropic GWB. In particular, we generalize their treatment to account for the 
modulation induced by the relative motion between the detector and the cosmic rest 
frame, and we highlight the role played by the time-dependent orientation and 
orbital motion of the LISA constellation. This analytic perspective  helps 
building intuition for the structure of the signal and will guide the numerical 
analysis presented in the subsequent sections.

\subsection{The GWB in a boosted frame}
We expand the gravitational wave metric perturbation in plane waves as
\begin{equation}
h_{a b}(\bx,t) = \int _{-\infty}^{\infty} \dd f \int \dd^2 \bn \sum_{\lambda} h_\lambda (f, \bn) e^\lambda_{a b} (\bn) e^{\ii 2\pi f (t - \bn\cdot \bx/c)},
\end{equation}
where $e^{\lambda}_{a b}$ with $\lambda = +,\times$ are the GW polarization tensors, normalized such that $e^{\lambda}_{a b} e_{\lambda'}^{a b} = 2 \delta^\lambda_{\lambda'}$. The reality condition
of the GW in coordinate space ensures $h_\lambda (f, \bn) = h^*_\lambda (-f, \bn)$. For a stationary unpolarized background, the 2-point function of the GW field satisfies 
\be\label{correlator}
\langle h_\lambda (f,  \bn) h^\star_{\lambda'} (f',  \bn')\rangle\,=\,
\frac{\delta_{\lambda \lambda'}}{2}\,\delta(f-f')\,\frac{\delta^{(2)} (\bn-\bn')}{4 \pi}\,I(f,\bn)\,,
\ee
where $\lambda = +,\times$. Since $I(f, \bn)$ is real, then $I(f, \bn) = I(-f, \bn)$, hence it is sufficient to consider the one-sided spectrum. 
We  assume that the GW intensity is isotropic in the rest frame of the GWB source, and we denote it as $\bar I(f)$. 
The relation with the dimensionless energy density is 
\be
\frac{\rho_{\rm GW}}{\rho_{\rm c}} = 4\pi^2/(3 H_0^2)\int_0^\infty f^2\bar{I}(f) \dd f\equiv \int \dd \log f\,\Omega_{\rm GW}(f)\,,
\ee
where in the last equality we introduced 
$\Omega_{\rm GW}(f)$, 
the dimensionless energy density per units of log frequency.
As is customary, throughout our analysis we consider the quantity $h^2 \Omega_{\rm GW}$,
which is independent of the uncertainty in the Hubble constant, conventionally written as
$H_0 = h\,100\,\mathrm{km\,s^{-1}\,Mpc^{-1}}$ and encoded in the parameter $h$.
For a flat spectrum, we denote the overall amplitude of $h^2 \Omega_{\rm GW}$
by $\Omega_*$.

Using the results of \cite{Cusin:2022cbb}, and since  $\bar I\propto \Omega_{\rm GW}/f^3$, kinematic effects induce Doppler anisotropies described by the relation
\be
\label{int_kin}
I(f, \bn)\,=\,{\cal D}\,\bar I\left({\cal D}^{-1}\,f \right)\,,\quad\text{with}\quad{\cal D}\,\equiv\,\frac{\sqrt{1-\beta^2}}{1- \bn \cdot {\bm \beta}}\,,
\ee
where ${\bm \beta}$ is our (relative) Doppler velocity whose norm is $\beta$. At linear order in $\beta$ this reduces to
\be\label{Idipolar}
\frac{I(f,\bn)}{\bar I(f)} \simeq 1 +\bn\cdot {\bm \beta}\, \left(1-n_I \right)\,,\qquad n_I\,\equiv\,\frac{\dd\,\ln \bar I}{\dd\,\ln f}\,,
\ee
where $n_I(f)$ is the tilt parameter, which is in general frequency dependent, unless the spectrum is a power law, and is related to the tilt parameter of $\Omega_{\rm GW}$ by $n_I(f) = n_t(f)-3$.

\subsection{The covariance matrix: general expressions}

We follow the derivation of the LISA covariance presented in \cite{Smith:2019wny}, 
and we extend it to the anisotropic case while accounting for the slow motion of the 
LISA constellation. To this end, we introduce a separation between a slow time 
variable $\tau$, describing the evolution of the detector position and orientation 
along its orbit, and a fast time variable $t$, associated with the variation of 
the phase measurements~\cite{Mentasti:2023uyi}.

We divide the data stream into time intervals of duration $T_{\rm Fourier}$, 
within which the detector configuration can be treated as approximately constant. 
These segments are labelled by discrete slow-time stamps $\tau_n$, and satisfy
\begin{equation}
\sum_{\tau_n} T_{\rm Fourier} = T \, ,
\end{equation}
where $T$ denotes the total observation time of the experiment.

The measured relative time difference at a vertex of the interferometer, $\Phi$, can be expressed in terms of the gravitational response in terms of an interferometer phase at that vertex, $S$, as well as the noise, $n$ as 
\be\label{Phi}
\Phi_{A_{BC}}(t;\tau_n)=S_{A_{BC}}(t;\tau_n)+n_{A_{BC}}(t)\,,
\ee
where the subscript ${A_{BC}}$ indicates the signal detected in the interferometer consisting of arms $AB$ and $AC$. 
Elsewhere in the literature, $A_{BC}$, $B_{CA}$, $C_{AB}$ are labeled as X, Y, Z,  which we sometime use for compact notation. We also define Fourier components with respect to  the fast time $t$ as
\be
\Phi_{A_{BC}}(t;\tau_n) = \int_{-\infty}^{\infty} \dd f {\rm e}^{2\pi \ii f t}\Phi_{A_{BC}}(f;\tau_n)\,\quad \Phi_{A_{BC}}(f;\tau_n) = \int_{-\infty}^\infty \dd t {\rm e}^{-2\pi \ii f t}\Phi_{A_{BC}}(t;\tau_n)\,,
\ee
with similar definitions for the intrinsic signal $S_{A_{BC}}$ and the additional noise $n_{A_{BC}}$. Even though the integrals on time should span an interval $T_{\rm Fourier}$, we approximate  this quantity  as  an infinite time span. Similarly, the integral on frequencies is    a sum over the discrete frequencies $f_p = p /T_{\rm Fourier}$ -- limited by the sampling frequency --  but we  take them as continuous approximation~\cite{Pitrou:2024scp}.

The relative time difference measured at  vertex ${A_{BC}}$ is (see e.g. \cite{Smith:2019wny})
\begin{equation}
S_{A_{BC}}(t;\tau_n) = \int_{-\infty}^\infty \dd f \int d^2 \bn \sum_\lambda {h}_\lambda(f,\bn)e^{i 2\pi f t} \Fresponse^\lambda_{A_{BC}}(\bn,f;\tau_n)\,,\label{eq:int_phase}
\end{equation}
with
\begin{align}
\Fresponse^\lambda_{A_{BC}}(\bn,f;\tau_n) &=  \frac{1}{2} e^{-i 2\pi f\bn \cdot \vec x_A(\tau_n)/c}e_{ab}^\lambda(\bn) \left[ \Fresponse^{ab}(\boldell_{AB}(\tau_n) , \bn,f) -\Fresponse^{ab}(\boldell_{AC}(\tau_n) , \bn,f)\right] \label{eq:gain1} \\
\Fresponse^{ab}(\boldell , \bn,f) &=
\frac{1}{2}W(f)\boldell^a \boldell^b \left( {\rm sinc}\left[\frac{f}{2 f_*}(1-\boldell\cdot\bn)\right] e^{-i\frac{f}{2f_*}(3+\boldell\cdot\bn)} \right. \\ \nonumber \quad\qquad & \left.  \qquad+ {\rm sinc}\left[\frac{f}{2 f_*}(1+\boldell\cdot\bn)\right] e^{-i\frac{f}{2f_*}(1+\boldell\cdot\bn)} \right) \,,
\label{eq:gain2}
\end{align}
gives the gain of a detector vertex. 
The vector $\boldell_{AB}(\tau)$ is a unitary vector which points from vertex $A$ to $B$, $f_*=c/(2\pi L)$, and $\bn$ is the direction of gravitational wave propagation.
LISA operates using measurement schemes based on time-delay interferometry (TDI)
techniques \cite{Tinto:2002de,Prince:2002hp}. When $W=1$, the expressions above describe the difference between the round-trip
light paths $ABA$ and $ACA$, corresponding to the Michelson configuration
(often referred to as the TDI 1.0 combination). In this limit the signal
effectively involves a single round trip of the laser beam along two arms of
the interferometer.
In practice, however, additional sources of noise -- most importantly laser
frequency noise -- require the use of more sophisticated TDI combinations (see e.g. \cite{Vallisneri:2005ji,Muratore:2020mdf,Hartwig:2021mzw}). 
These configurations construct synthetic interferometric observables by
combining phase measurements along longer light paths, in which the laser
signals propagate through multiple LISA vertices, before being recombined.
The phase accumulated along these extended paths can be described by the same
expressions introduced above, but evaluated at appropriately time-shifted
arguments. In particular, additional round trips correspond to time offsets of
the form $t \rightarrow t - 2L/c$. In the Fourier domain, these time delays
translate into phase factors in the time-series transform.
For example, choosing
\begin{equation}
W(f) = 1 - e^{-2 i f/f_*}
\end{equation}
accounts for the interference between signals separated by a round-trip light
travel time across the LISA arm length. This configuration effectively captures
the response associated with a full light circulation in the triangular LISA
constellation, and corresponds to what is often referred to as the TDI 1.5
combination.
We refer the reader to \cite{Tinto:2020fcc} for a comprehensive review of TDI techniques,
and their implementation in a LISA-like configuration.

The correlation of the measurment is defined from the statistical average
\be
\langle \Phi_X (t;\tau_n) \Phi_{X'}(t';\tau_{n'})\rangle = \delta_{nn'} C_{XX'}(t-t';\tau_n) = \frac{1}{2}\delta_{nn'}\int_{-\infty}^{\infty} \dd f {\rm e}^{2\pi \ii f (t-t')} {C}_{XX'}(f;\tau_n)\,,
\ee
where the subscripts $X, X'$ denote any one of the vertices $A_{BC}, B_{CA}, C_{AB}$. One can equivalently work directly in Fourier space and define 
\be
\langle \Phi_X (f;\tau_n) \Phi^\star_{X'}(f';\tau_{n'})\rangle = \frac{1}{2}\delta_{nn'}\delta(f-f'){C}_{XX'}(f;\tau_n)=\langle \Phi_a \Phi^\star_{a'}\rangle =  {C}_{aa'}\,,
\ee
where in the last expressions we introduced a compact index notation $a = (X,f,\tau_n)$. Note that the covariance matrix is Hermitian, that is $C_{aa'} = C^\star_{a'a}$, hence ${C}_{XX'}(f;\tau_n) = {C}^\star_{X'X}(f;\tau_n)$ and ${C}_{XX'}(t-t';\tau_n) = {C}^\star_{X'X}(t'-t;\tau_n)$. 
This correlation is the sum of the  signal contribution and the noise contribution. 
The signal part receives a contribution from the isotropic part of the GW background, and a part due to its dipolar modulation. Eventually, the general form of the correlation function spectral density is
\be\label{CgeneralXX}
{C}_{XX'}(f;\tau)={S}_{XX'}(f)+ {N}_{XX'}(f) = {M}_{XX'}(f,\tau)  + {D}_{XX'}(f,\tau)
+ {N}_{XX'}(f)\,,
\ee
where the first term is the monopolar contribution, the second is the dipolar modulation due to the velocity with respect to the GW background frame, and the last contribution is the noise. 

The monopolar contribution is related to the antenna pattern functions through
\be \label{covF}
{M}_{XX'}(f;\tau_n)=\int\frac{\dd^2\bn}{4\pi} \sum_{\lambda=+,\times} \Fresponse^\lambda_X(\bn,f;\tau_n) \Fresponse^{\lambda \star}_{X'}(\bn,f;\tau_n)I(f,\bn)\,,
\ee
and the dipolar contribution is given in section~\ref{sec:dipolarresponse}. The stationary noise contribution is defined by
\begin{equation}
\langle n_X (f;\tau_n) n^\star_{X'}(f';\tau_{n'})\rangle = \frac{1}{2}\delta_{nn'}\delta(f-f'){N}_{XX'}(f)\,.
\end{equation}

\subsection{LISA orientation}
 
LISA, in good approximation, can be assumed as   an equilateral triangle of fixed side $L$, with a varying orientation. Such an approximation is valid in the limit that the LISA satellites are on small eccentricity orbits~\cite{Rubbo:2003ap}. In order to set its orientation we  need to specify $\boldell_{AB}(\tau)$ and $\boldell_{AC}(\tau)$, since $\boldell_{BC}(\tau)=\boldell_{AC}(\tau)-\boldell_{AB}(\tau)$. We fix ${\bm e}_z$ as the unit vector normal to the ecliptic plane, and we define the unit vectors in the ecliptic plane
\be\label{RefDirections}
\boldell^o_{AB}= -\boldell^o_{BA} = {\bm x}^o_B - {\bm x}^o_A =  (1/2, \sqrt{3}/2, 0)\,,\qquad \boldell^o_{AC}=-\boldell^o_{CA}={\bm x}^o_C - {\bm x}^o_A = (-1/2, \sqrt{3}/2, 0)\,.
\ee
The general orientation is obtained by a
rotation controlled by a matrix $R^{\rm LISA}$ acting on these arm directions (see appendix~\ref{App:RLISA}):
\be\label{rotLISA}
R^{\rm LISA}(\tau) \equiv R_z(\Omega \tau)\cdot R_y(\pi/3)\cdot R_z(-\Omega \tau+\sigma)=R_{[R_z(\Omega \tau)\cdot {\bm e}_y]}(\pi/3)\cdot R_z(\sigma)\,,
\ee
where $\sigma$ is a constant phase associated with the triangle orientation at initial time, and $\Omega$ is the year frequency. $R_{[{\bm a}]}(\theta)$ denotes rotation around an axis ${\bm a}$ by angle $\theta$, which we abbreviate as $R_z$ when the axis is ${\bm e}_z$ and similarly for $R_y$. At any time, the orientation is given by
\be
\boldell_{AB}(\tau) = R^{\rm LISA}(\tau) \cdot \boldell^o_{AB}\,,\quad \boldell_{AC}(\tau) = R^{\rm LISA}(\tau) \cdot \boldell^o_{AC}\,.
\ee
For a representative graphical
illustration, see e.g. Fig.~(6.3) of~\cite{LISA:2024hlh}.

\subsection{Response function to the isotropic
background}

The monopole (isotropic) contribution to the correlation spectral density can be expressed in terms of a response function defined by
\begin{equation}\label{EqMRI}
{M}_{XX'}(f;\tau_n) = {\cal R}_{XX'}(f;\tau_n) \, \bar{I}(f) \, .
\end{equation}
Using Eq.~(\ref{covF}) for an isotropic intensity distribution, we obtain the explicit expression
\begin{equation}
{\cal R}_{XX'}(f;\tau_n) = \int \frac{\dd^2 \bn}{4\pi} 
\sum_{\lambda=+,\times} 
\Fresponse^\lambda_X(\bn,f;\tau_n)
\Fresponse^{\lambda \star}_{X'}(\bn,f;\tau_n)\, .
\label{eqn:rint}
\end{equation}

To evaluate this expression using Eq.~\eqref{eq:gain1}, we first rewrite the polarization sum
$\sum_{\lambda=+,\times} \epsilon^\lambda_{ij}\epsilon^\lambda_{kl}$
in terms of products of the direction vector $n^i$ and the identity tensor $\delta^{ij}$. 
We then expand the response in powers of the dimensionless frequency parameter $f$, which generates angular structures involving products of $n^i$. 
The resulting angular integrals can be evaluated using standard identities for averages over the sphere, such as those reported in Eqs.~(2.3) of~\cite{Thorne:1980ru}.

After performing these angular integrations, we find that the response function does not depend explicitly on the slow time $\tau_n$. It can therefore be written as
\begin{equation}
{\cal R}_{XX'}(f) = \int \frac{\dd^2 \bn}{4\pi} 
\sum_{\lambda=+,\times} 
\Fresponse^{o,\lambda}_X(\bn,f)
\Fresponse^{o,\lambda \star}_{X'}(\bn,f)\, ,
\label{eqn:rintbetter}
\end{equation}
where $\Fresponse^{o,\lambda}_X(\bn,f)$ are defined as in Eq.~\eqref{eq:gain1}, but using the fixed reference vectors introduced in Eq.~\eqref{RefDirections}.

The resulting response matrix has the general structure
\begin{equation}\label{response0}
 \left( {\cal R}_{XX'}(f)\right)=
\left(
\begin{array}{ccc}
{\mathcal{R}}_1& {\mathcal{R}}_2&{\mathcal{R}}_2\\
{\mathcal{R}}_2&{\mathcal{R}}_1&{\mathcal{R}}_2\\
{\mathcal{R}}_2&{\mathcal{R}}_2&{\mathcal{R}}_1
\end{array}
\right)\, .
\end{equation}
The functions ${\mathcal{R}}_1$ and ${\mathcal{R}}_2$ admit a series expansion in the dimensionless parameter $x=f/f_*$. Their leading terms are
\begin{align}\label{R1}
{\mathcal{R}}_1(f)&=|W|^2\left(\frac{3}{10}-\frac{169}{1680}x^2+\frac{85}{6048}x^4-\frac{165073}{159667200}x^6+\frac{132439}{2830464000}x^8 +\mathcal{O}(x^{10}) \right)\,,\\
{\mathcal{R}}_2(f)&=|W|^2\left(-\frac{3}{20}+\frac{169}{3360}x^2-\frac{85}{12096}x^4+\frac{29239}{45619200}x^6-\frac{251389}{5660928000}x^8 +\mathcal{O}(x^{10}) \right)\, .
\label{R2}
\end{align}
These expressions agree with the results of \cite{Smith:2019wny}.
\subsection{Dipolar response function}\label{sec:dipolarresponse}

Similarly, substituting Eq.~(\ref{Idipolar}) into Eq.~(\ref{covF}) and retaining the
terms linear in the peculiar velocity, we find that the dipolar contribution to
the correlation spectral density can be expressed in terms of vector-valued
response functions $D^j_{XX'}(f)$ as
\begin{equation}\label{DefDXXi}
{D}_{XX'}(f;\tau_n)  = \beta^i \, R^{\rm LISA}_{ij}(\tau_n) \, D^j_{XX'}(f)
\,(1-n_I(f))\,\bar{I}(f) \, .
\end{equation}
where the matrix $R^{\rm LISA}_{ij}$ is introduced in Eq.~\eqref{rotLISA}.  
The explicit form for each component of the dipole response function is obtained from\footnote{For simplicity we neglect the additional yearly modulation of the
dipole induced by the orbital motion of the detector around the Sun. This
modulation affects the construction of the optimal estimator, but not its
variance, and therefore does not impact the error estimates presented in our
analysis.}
\begin{equation}
D^i_{XX'}(f) = \int \frac{\dd^2 \bn}{4\pi}
\sum_{\lambda=+,\times}
n^i\,\Fresponse^{o,\lambda}_X(\bn,f)
\Fresponse^{o,\lambda \star}_{X'}(\bn,f)\, .
\label{resp_dip}
\end{equation}
The integral in Eq.~\eqref{resp_dip} can be evaluated numerically. However, its structure is  constrained by symmetry considerations, which allow us to anticipate its general form. First, the presence of the factor $n^i$ implies that $D^i_{XX'}$ transforms as a vector under spatial rotations. Second, the product of response functions $\Fresponse_X \Fresponse_{X'}^\star$ depends only on the geometry of the interferometer arms, and in particular on the relative orientations of the links connecting the spacecraft.
Combining these observations with the discrete symmetries of the LISA constellation -- in particular its invariance under cyclic permutations of the spacecraft -- we can anticipate that the dipolar response must be aligned along the directions of the interferometer arms. The result takes indeed the form
\begin{align}\label{RBeta}
\left(D^i_{XX'}(f)\right)=
\ii\, {\mathcal{R}}_3(f)
\left(
\begin{array}{ccc}
0& \ell^{o,i}_{AB}&\ell^{o,i}_{AC}\\
-\ell^{o,i}_{AB}&0&\ell^{o,i}_{BC}\\
-\ell^{o,i}_{AC}&-\ell^{o,i}_{BC}&0
\end{array}
\right)\quad \Leftrightarrow\quad D^i_{XX'}(f) = \ii {\cal R}_3(f) \ell^{o,i}_{X X'}\,,
\end{align}
where $\ell^{o,i}_{XY}$ denotes the unit vector along the arm connecting spacecraft $X$ and $Y$. The correlation matrix is purely imaginary and antisymmetric, consistently with the Hermiticity of the full correlation matrix.  
The function ${\mathcal{R}}_3(f)$ depends only on the dimensionless frequency $x=f/f_\star$. It admits the following expansion at low frequency:
\begin{equation}
\label{exp_R3}
{\mathcal{R}}_3(f) =
|W|^2
\left(
\frac{1}{336}x^3
-\frac{11}{13440}x^5
+\frac{673}{6652800}x^7
+ {\cal O}(x^9)
\right)\, .
\end{equation}
We represent in  Fig.~\ref{fig:response_functionA}
the full numerical evaluation of ${\mathcal{R}}_3(f)$.
The components of the dipolar response in Eq.~\eqref{RBeta} depend on the orientations of the detector arms. When contracted with the velocity components as in Eq.~\eqref{DefDXXi}, only the components of the velocity vector lying in the LISA plane yield a non-vanishing contribution. Moreover, from the properties of the integrand in Eq.~\eqref{resp_dip}, we find that the dipolar response vanishes in the limit $f/f_\star \to 0$, as shown in Eq.~\eqref{exp_R3}.\footnote{In the presence of parity violation, the background can be circularly polarized. The dipolar contribution~\eqref{DefDXXi} is then supplemented by a term that is detailed in appendix~\ref{SecPolarization}, and which is sensitive to the components of the dipole in the direction orthogonal to the LISA plane.\cite{Domcke:2019zls}.}
In order to compute the quantity $R^{\rm LISA}_{ij}(\tau_n) D^j_{XX'}(f)$
appearing in Eq.~\eqref{DefDXXi}, it is sufficient to replace the reference arm
vectors according to
\be
\ell^{o,i}_{XY} \;\rightarrow\; \ell^i_{XY}(\tau_n)
\ee
in Eq.~\eqref{RBeta}. 
The fact that a unique function of frequency factorizes out
in the  response to the dipole
can be understood more generally  through a spherical harmonic analysis of LISA response functions, see section 4 of \cite{LISACosmologyWorkingGroup:2022kbp} for details. 
As we show in Appendix~\ref{quadrupole}, the situation for the quadrupole is qualitatively different. In that case, the response is described by rank-2 tensors, and two independent frequency-dependent functions are required to parametrize the signal. This leads to a richer geometrical structure, involving both diagonal and off-diagonal components of the correlation matrix. However, despite this increased complexity, the underlying symmetry arguments remain the same and provide a powerful guide to the final result. The analysis presented above highlights a key feature of LISA response to anisotropies: its structure is fully fixed by symmetry, and depends on a single scalar function of frequency multiplying simple geometrical vectors.

\subsection{The noise correlation}\label{Noise}Schematically, the residual noise components entering each TDI channel can be grouped
into two effective contributions: the ``Optical Metrology System'' (OMS) noise, which
includes, for example, shot noise in the interferometric readout, and the
``acceleration'' noise associated with random displacements of the proof masses,
caused for instance by local environmental disturbances.

Our current understanding of the LISA noise budget is largely based on the results
of the LISA Pathfinder experiment \cite{Armano:2016bkm}, complemented by laboratory
tests. The power spectral densities of the OMS and acceleration noise contributions
to the relative length measurements are given by (see, e.g., 
Eq.~(2.2) of \cite{Flauger:2020qyi}, Eqs.~(9--13) of \cite{Babak:2021mhe}, or
Eqs.~(10) and (11) of \cite{Robson:2018ifk})
\begin{align}\label{NoiseP}
P_{\text{OMS}}(f; P)&=P^2  \frac{1}{\text{Hz}}\left[1+\left(\frac{2 \text{mHz}}{f}\right)^4\right]\left(\frac{\text{pm}}{L}\right)^2\,,\\
P_{\text{acc}}(f; A)&=A^2  \frac{1}{\text{Hz}}\left[1+\left(\frac{0.4 \text{mHz}}{f}\right)^2\right]\left[1+\left(\frac{f}{8 \text{mHz}}\right)^4\right]\left(\frac{\text{Hz}}{2\pi f}\right)^4\left(\frac{\text{fm}}{L}\right)^2\,.
\end{align}
ESA mission specifications require the amplitudes to be $P=15$ and $A=3$, with
$\pm 20\%$ margins. We refer to this noise model as \emph{fiducial} LISA.
We will also
consider a future instrument operating in the same frequency band as LISA,
with both $A$ and $P$ reduced by a factor of $10$, which we refer to as \emph{futuristic} LISA. Note that this corresponds to a rather unrealistic noise improvement. However, as we will explain it reproduces the same result one would obtain with an instrument which improves the noise by one order of magnitude and  increases the nominal mission time with respect to LISA.

For the OMS contribution, we adopt the simplifying assumption that the noise
spectra of all links are identical, stationary, and uncorrelated. Similarly,
for the acceleration noise we assume that the fluctuations of the proof masses
are isotropic and stationary, that all test masses have identical power
spectra, and that fluctuations of different masses are uncorrelated.
Furthermore, we assume that the three spacecraft form an equilateral triangle
with arm length $L = 2.5 \times 10^9\,\mathrm{m}$. Under these assumptions,
the total noise power spectral density takes the form
\be\label{noise}
\mathcal{N}\equiv \left(  N_{A_{BC} X_{YZ}}\right)=
\left(
\begin{array}{ccc}
N_1& N_2&N_2\\
N_2&N_1&N_2\\
N_2&N_2&N_1\\
\end{array}
\right)\,,
\ee
where
\begin{align}
N_1(f; A, P) &= |W|^2 \left\{ \left[3 + \cos\!\left(\frac{2f}{f_*}\right)\right]
P_{\text{acc}}(f, A) + P_{\text{OMS}}(f, P)\right\}\,,\\
N_2(f; A, P) &= -\frac{1}{2} |W|^2 \cos\!\left(\frac{f}{f_*}\right)
\left[4 P_{\text{acc}}(f, A) + P_{\text{OMS}}(f, P)\right]\,,
\end{align}
and $|W|^2 = 4 \sin^2\!\left(f/f_*\right)$. 
Our final expressions for the noise agree with \cite{Robson:2018ifk}, but are
smaller by a factor of $4$ compared with Eqs.~(A32--A33) of
\cite{Flauger:2020qyi}, Eq.~(56) of \cite{Cornish:2001bb}, and Eq.~(19) of
\cite{Babak:2021mhe}. This difference arises because our signal is normalized
by $2L$, rather than by $L$.

\subsection{Standard diagonalisation: AET channels}
In the literature it is usually assumed that the intensity is direction independent. In this case, because the monopole response function
(\ref{response0}) and the noise covariance (\ref{noise}) share the same
matrix structure, it is possible to construct a basis of three orthogonal
(i.e., statistically independent) linear combinations of the signals in
which both the monopole response and the noise covariance become diagonal.
It is customary to introduce combinations of the form
\be\label{defPassagematrix}
\Phi_O = \Phi_X\,{\cal P}_{XO}\,,
\ee
where $O=\{A,E,T\}$ and $X=\{A_{BC},B_{CA},C_{AB}\}$. The transformation
matrix is
\be
\left(\mathcal{P}_{XO}\right)=\left(
    \begin{array}{ccc}
    1/\sqrt{6}&   1/\sqrt{2}&   1/\sqrt{3}\\
  -2/\sqrt{6}&0&1/\sqrt{3}\\
    1/\sqrt{6}&-1/\sqrt{2}&1/\sqrt{3}
     \end{array}
    \right)\,.
\ee
In the $\{A,E,T\}$ basis both the response matrix and the noise covariance
become diagonal,
\be\label{CAET}
{\cal R}_{AET}=\mathcal{P}^T {\cal R}\mathcal{P}\,,\qquad
{N}_{AET}=\mathcal{P}^T {N}\mathcal{P}\,,
\ee
where ${\cal R}_{AET}$ is diagonal with entries ${\cal R}_{A}$,
${\cal R}_{E}$ and ${\cal R}_{T}$, and similarly for the noise matrix
${N}_{AET}$. 
The eigenvalues are related to the elements of the original matrix through
\begin{equation}
{\cal R}_{A}={\cal R}_E={\cal R}_1-{\cal R}_2\,,\qquad
{\cal R}_T={\cal R}_1+2{\cal R}_2\,,
\end{equation}
with analogous relations holding for the noise eigenvalues.
\begin{figure}[t!]
\centering
\includegraphics[width=0.48\textwidth]{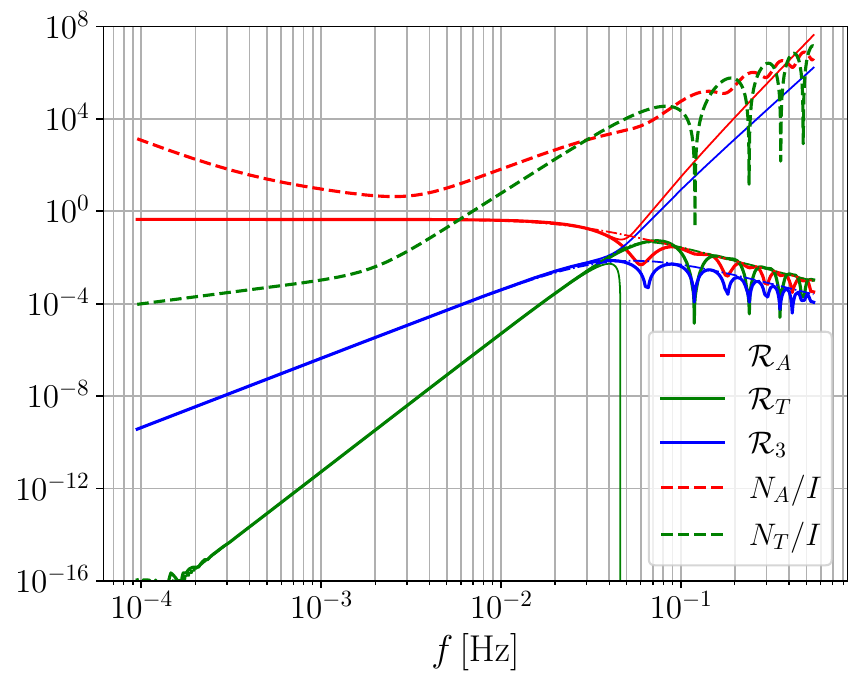}\quad
\includegraphics[width=0.48\textwidth]{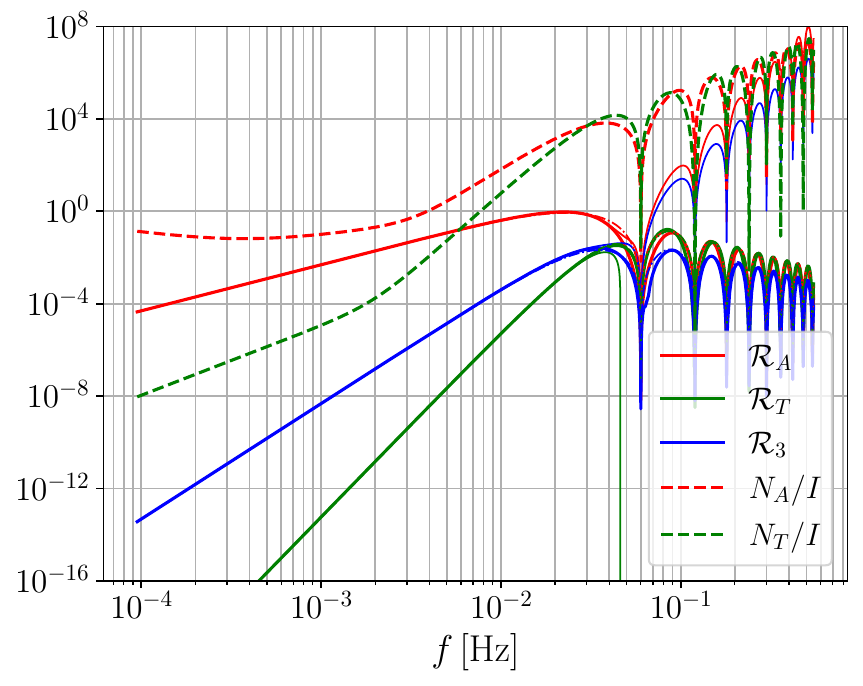}
\caption{Left panel:
Response functions $\mathcal{R}_A$, $\mathcal{R}_T$ and $\mathcal{R}_3$
for $W=1$.
Right panel:
Response functions for $|W|^2 = 4 \sin^2(f/f_\star)$.
The small-frequency series expansions are shown as thin solid lines
(they overlap with the thick curves at low frequency), while the dashed
lines represent the noise contributions $N_A$ and $N_T$ -- discussed in
Sec.~\ref{Noise} -- expressed in units of $I(f)\propto f^{-3}$ assuming
$\Omega_\star=10^{-12}$ for a scale-free spectrum. The dashed-dotted lines are the fitting formulas in Eqs. \eqref{fitting1}, \eqref{fitting2} and \eqref{fitting3}.}
\label{fig:response_functionA}
\end{figure}

Using Eqs.~(\ref{R1}) and (\ref{R2}), we find that the $A$ and $E$ response
functions are constant for $f<f_*$ and scale as $f^{-2}$ for $f>f_*$. In
contrast, the $T$ response function scales as $f^{6}$ for $f<f_*$ and as
$f^{-2}$ for $f>f_*$.
 A good fit for the response functions is given by 
\begin{align}\label{fitting1}
{\cal R}_A^{\text{Fit}} &= \frac{9}{20}|W|^2 \left[1+\frac{9}{16}\left(\frac{ f}{f_*}\right)^2\right]^{-1}\,,\\
\label{fitting2}
{\cal R}_T^{\text{Fit}} &=\frac{1}{4032}\left(\frac{f}{f_*}\right)^6 |W|^2 \left( 1+\frac{61}{360}\left[\frac{f}{f_*}\right]^2+\frac{5}{16128}\left[\frac{f}{f_*}\right]^8\right)^{-1}\,.
\end{align}
The $T$ channel is significantly less sensitive to the signal than the $A$
and $E$ channels. For this reason it is commonly used as a monitor of the
instrumental noise.

Using the results of Sec.~\ref{Noise}, the low-frequency expansion of the noise eigenvalues takes the form 
\begin{align}
N_A=N_E &\simeq \frac{3}{2}|W|^2\left(4 P_{\text{acc}}+P_{\text{OMS}}\right)\,,\\
N_T &\simeq \frac{1}{2} |W|^2\left(\frac{f}{f_*}\right)^2\left[
\left(\frac{f}{f_*}\right)^2 P_{\text{acc}}+ P_{\text{OMS}}\right].
\end{align}
In this regime, $P_{\rm acc} \propto f^{-6}$ and $P_{\rm OMS} \propto f^{-4}$. Consequently, the two single-link noise contributions enter $N_T$ at the same order, implying,  if we take $|W|^2 \propto f^2$ into account, $N_T \propto f^{0}$. In contrast, for the noise in the $A/E$ channels, $P_{\rm acc}$ dominates in the low-frequency limit, leading to $N_A \propto f^{-4}$ (again including $|W|^2 \propto f^2$).

The structure of the dipolar contribution to the correlation matrix is
quite different from the one  of the monopole response, or the noise. In
particular, it is not diagonal in the $\{A,E,T\}$, basis and it remains purely
imaginary and antisymmetric. Nevertheless, the change of basis leads to a
considerably simpler form. To this end, we define
$\tilde\ell^{o}_{AC} \equiv R_z(\pi/2)\cdot \ell^{o}_{AC}$, such that
$(\ell^{o}_{AC},\tilde\ell^{o}_{AC})$ forms an orthonormal basis in the
LISA plane. With the transformation~\eqref{defPassagematrix}, the dipolar
response matrix becomes
\be\label{MagicPDP}
(\mathcal P_{XO})^{T}\, D^i(f)\,(\mathcal P_{XO})
=
\begin{pmatrix}
0 & \dfrac{D^{i}_{AB}+D^{i}_{BC}+D^{i}_{CA}}{\sqrt{3}}
& \dfrac{D^{i}_{AB}-D^{i}_{BC}}{\sqrt{2}}\\[10pt]
-\dfrac{D^{i}_{AB}+D^{i}_{BC}+D^{i}_{CA}}{\sqrt{3}}
& 0
& \dfrac{D^{i}_{AB}+D^{i}_{BC}-2D^{i}_{AC}}{\sqrt{6}}\\[10pt]
-\dfrac{D^{i}_{AB}-D^{i}_{BC}}{\sqrt{2}}
& -\dfrac{D^{i}_{AB}+D^{i}_{BC}-2D^{i}_{CA}}{\sqrt{6}}
& 0
\end{pmatrix}.
\ee
Using Eq.~(\ref{RBeta}), this expression can be written in the more
transparent form 
\begin{equation}\label{RBetaAET}
\left( D^i_{O O'}(f)\right)=\ii {\mathcal{R}}_3(f)\sqrt{\frac{3}{2}}\left(
\begin{array}{ccc}
0& 0&\tilde\ell^{o,i}_{AC} \\
0& 0&\ell^{o,i}_{AC}\\
-\tilde\ell^{o,i}_{AC}  &-\ell^{o,i}_{AC}&0\\
\end{array}
\right)\,,
\end{equation}
with $O, O'=(A, E, T)$. The $AE$ component vanishes because LISA is assumed
to form a perfect equilateral triangle. Consequently, the dipole induces
off-diagonal correlations only between the $AT$ and $ET$ channels, but not
between $AE$.

We find the following convenient fitting formula for the dipole response function
$\mathcal{R}_3(f)$:
\be\label{fitting3}
{\cal R}^{\mathrm{Fit}}_3(f)=\frac{1}{336} \left(
\frac{ f}{ f_\star}\right)^3
\left(1+\frac{9}{50} \left( \frac{ f}{ f_\star}\right)^2
\right)^{-\frac{5}{2}}\,.
\ee
In Fig.~\ref{fig:response_functionA} we plot the response functions
${\mathcal{R}}_A$, ${\mathcal{R}}_T$, and ${\mathcal{R}}_3$, together with
their low-frequency expansions.

\subsection{Monopole detectability}

Before discussing the dipole contribution, we pause to present a
consistency check by computing the signal-to-noise ratio (SNR) for the amplitude of the isotropic component of the signal. We find\footnote{The SNR is found by starting from the covariance matrix (\ref{CgeneralXX}) and computing the Fisher matrix element associated with the monopole amplitude, i.e. $\text{SNR}_{\Omega}^2=\{ \partial_{\Omega}C; \partial_{\Omega}C\}$, where $\Omega$ is a bookkeeping parameter associated with the monopole amplitude that is set to one after taking the derivative. The brackets denote the inner product; see \ref{inner}.}
\be\label{DefSNR2Omega}
\text{SNR}^2_{\Omega} = T \int_0^\infty \dd f\, {\cal F}_M(f)\,,
\ee
with 
\be\label{DefSNR2Omega2}
{\cal F}_M(f) =
\left[
2\left(\frac{{\cal R}_A \bar{I}(f)}{{\cal R}_A \bar{I}(f) + N_A(f)}\right)^2
+
\left(\frac{{\cal R}_T \bar{I}(f)}{{\cal R}_T \bar{I}(f) + N_T(f)}\right)^2
\right] .
\ee
In practice, the frequency integral is performed over the range
$f_{\rm min}=10^{-4}\,{\rm Hz}$ to $f_{\rm max}=0.1\,{\rm Hz}$. For a perfect experiment we obtain $\text{SNR}^2_{\Omega} \simeq 3 T f_{\rm max} $ which we interpret as the fact that we have $2 T f_{\rm max}$ independent frequency modes, each with three channels, each of which with a contribution $\text{SNR}^2 = 1/2$ (see e.g. section VII.C.1 of~\cite{Pitrou:2024scp}). For an observation time $T=4$  years of observation $3T f_{\rm max} \simeq 3.7\times 10^7 $ this means that $\text{SNR} \lesssim 6000$.

If we  consider the effect of noise in the experiment,  for a scale-invariant spectrum we obtain $\text{SNR}=10$ when
$h^2\Omega(f_\star)=1.1\times 10^{-13}$, and $\text{SNR}=5$ when
$h^2\Omega(f_\star)=5.8\times 10^{-14}$. 
For comparison, \cite{Smith:2019wny} finds a sensitivity curve larger by
a factor of $4$. This difference originates from the normalization of
the noise: in their definition the variable corresponds to the relative
time delay for a round trip, and is therefore divided by $2L$, whereas
the noise used in their analysis corresponds to the sum of the
individual time delays of each one-way trip, which is normalized by $L$.

\begin{figure}[ht!]
\centering
\includegraphics[width=0.49\textwidth]{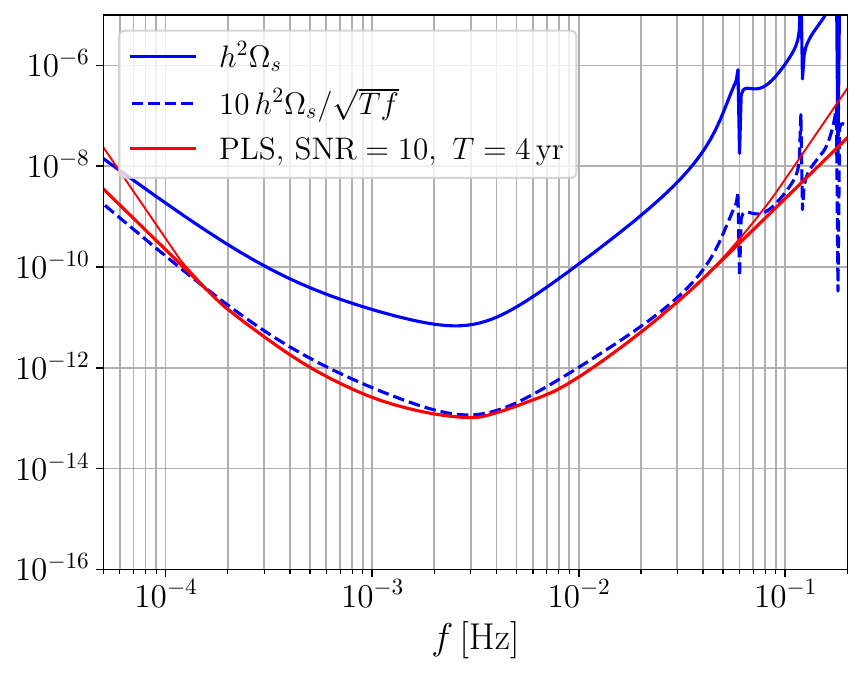}
\includegraphics[width=0.49\textwidth]{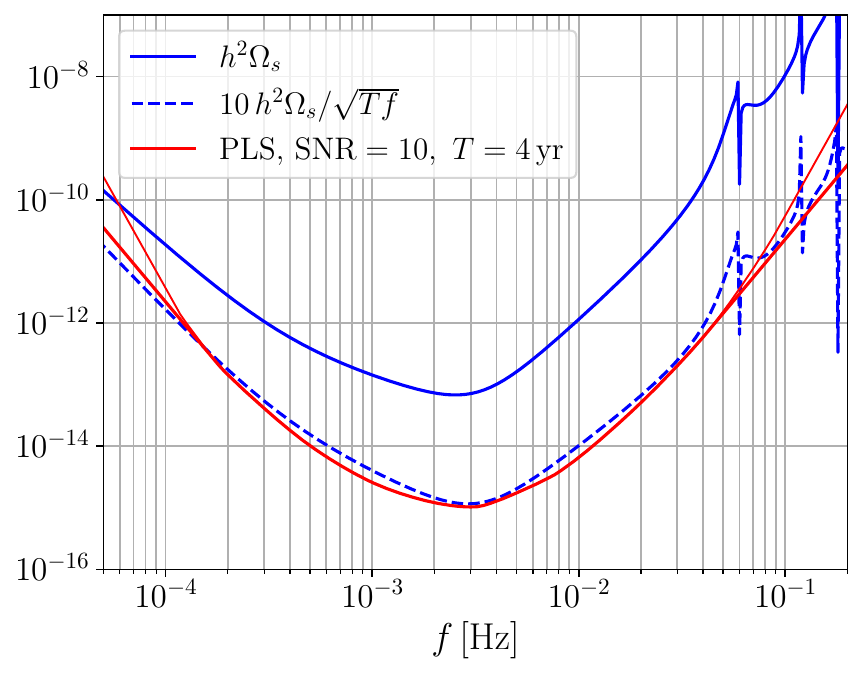}
\caption{Sensitivity curves for the GW monopole.
The blue lines correspond to the estimates based on~\eqref{NaiveOmegas}.
The PLS, described in
\cite{Thrane:2013oya,Caprini:2019pxz}, is shown in red.
The PLS depends on the range of spectral indices used to construct the
envelope: the thick red curve corresponds to the $\Omega_{\mathrm{GW}}$ spectral index given by $|n_t|<4$,
whereas the thin red curve is obtained assuming $|n_t|<6$. 
The left panel corresponds to the mission requirements $A=3$ and $P=15$,
while in the right panel these parameters are reduced by a factor of $10$.
}
\label{fig:Omegas}
\end{figure}

In addition, we can construct a rough estimate of the sensitivity curve
following~\cite{Caprini:2019pxz}. We first define
\be\label{NaiveOmegas}
\Omega_s(f) \equiv
\left(\frac{4\pi^2 f^3}{3 H_0^2}\right)
\left(
2\frac{{\cal R}_A^2}{N_A^2}
+
\frac{{\cal R}_T^2}{N_T^2}
\right)^{-1/2},
\ee
such that an estimate of the sensitivity at $\text{SNR}=10$ for an
observing time $T$ is given by $10\,\Omega_s(f)/\sqrt{T f}$. 
This estimate can be refined following~\cite{Thrane:2013oya} by
constructing the envelope of all power-law spectra that satisfy a given
SNR threshold, i.e. the so-called \emph{power-law sensitivity curve} (PLS).
These estimates of the LISA sensitivity are shown in
Fig.~\ref{fig:Omegas}. All curves lie slightly below those of
Fig.~2 in~\cite{Caprini:2019pxz}, primarily because we assume an
observing time $T=4\,{\rm yr}$ (instead of $T=3\,{\rm yr}$) and we combine
the signal from the three $A$, $E$, and $T$ channels rather than using a
single channel. Note that, once the signal amplitude cannot be neglected with respect to the noise, the former must be carefully included in the denominator of SNR estimate.

\section{Building an estimator for the velocity}
\label{estimator}
The goal of this section is to construct an estimator for the components
of the kinematic dipole. We develop a pedagogical derivation and present
compact analytic expressions for the variance of the velocity
components. In particular, we explicitly show that LISA cannot
reconstruct the velocity component orthogonal to its plane if the
detector was static. The reconstruction of the full velocity vector is therefore only possible by exploiting the motion of LISA around the ecliptic.

\subsection{Inner product on matrix space}\label{inner}

To construct the optimal quadratic estimator we introduce the following
mathematical tools. If $C$ is a positive-definite Hermitian matrix, we
define an inner product in matrix space as 
\be \label{innerproduct}
\{ A;B\}_C \equiv \frac{1}{2}{\rm Tr}(A \cdot C^{-1} \cdot B \cdot C^{-1})
= \frac{1}{2}A_{ab}C^{-1}_{bc} C^{-1}_{da}B_{cd}\,.
\ee
In general, the Fisher matrix fora set of model parameters index by $i$ is defined by
\be
{}^C {F}_{ij} = \{ \partial_{i} C; \partial_{j} C \}_{C}\,.
\ee
To build an estimator for $\beta^i$, the dipole velocity components in the ecliptic coordinates, we follow the approach of
\cite{Bond:1998zw}.

\subsection{A static LISA}

In order to emphasize the importance of the motion
of LISA, we first consider a fictitious situation in which LISA remains
static. In such a  case the dependence on the slow time $\tau_n$ can be
ignored, and we may set $R^{\rm LISA}_{ij}=\delta_{ij}$.

The Fisher matrix when constraining the dipole components is defined in general as
\be
{}^C {F}_{\beta_i \beta_j} = \{ \partial_{\beta_i} C; \partial_{\beta_j} C \}_{C}\,.
\ee
Since we have
\be
\partial_{\beta_i} C_{aa'}  =
\tilde{D}^i_{aa'}\,,
\qquad
\tilde{D}^i_{aa'} \equiv
D^i_{aa'}(1-n_I(f))\bar{I}(f)\,,
\ee
where we use the generalized indices $a=(O,f,\tau_n)$, the Fisher matrix becomes
\be
{}^C {F}_{\beta_i \beta_j} = \{\tilde D^i;\tilde D^j\}_C\,.
\ee
When evaluating the inner products,
a well motivated approximation consists in replacing the inverse of the full covariance matrix
$C^{-1}$ with the combination $(M+N)^{-1}$.
In fact,
the neglected dipole contribution to the inverse covariance gives corrections only at higher order in $\beta$.\footnote{
We have verified numerically that using the full $C^{-1}$ in the inner product yields results fully consistent with those obtained analytically and presented throughout this work.} 
We therefore consider
\be
{}^{M+N} {F}_{\beta_i \beta_j} =\{\tilde D^i;\tilde D^j\}_{M+N}\,.
\ee
The quadratic estimator for the velocity is
\be
V^i = ({}^C{F}^{-1})_{\beta_i \beta_j}\, {}^C X^j\,,
\qquad
{}^C X^j \equiv\{ \tilde D^j;
\Phi_a \Phi^\dagger_{b} - N_{ab}-M_{a b} \}_C\,.
\ee
The estimator is unbiased since
\be
\langle {}^C X^j \rangle
= \{ \tilde D^j;
\langle S_a S^\dagger_b \rangle -M_{ab}\}_C
= \beta^i
\{ \tilde D^j; \tilde D^i \}_C
= {}^C {F}_{\beta_j \beta_i} \beta^i\,,
\ee
where the signal $S_a$ is defined in eq.\,(\ref{Phi}). The minimum variance is achieved when the full covariance $C$ is used in
the inner products defining the estimator. In that case one finds, after
some algebra,
\be
\langle V^i V^j \rangle = {}^C{F}^{-1}_{\beta_i \beta_j}\,,
\ee
as expected. The squared error on each velocity component is therefore
given by the corresponding diagonal element of the inverse Fisher
matrix.

However, this procedure cannot be applied directly in the static LISA
case. Indeed one finds
\be\label{sumAO}
F^{\rm st.}_{\beta_i \beta_j} \propto
\sum_{O O'}
D^i_{OO'}D^j_{O'O}
=
\sum_{O=A,E}2 D^i_{OT}D^j_{TO}
=
3 (\mathcal{R}_3(f))^2\left(\delta^{ij}-e_z^i e_z^j\right)\,,
\ee
so that the Fisher matrix is not invertible. Its general structure is
\be
F^{\rm st.}_{\beta_i \beta_j} =
\left(\delta^{ij}-e_z^i e_z^j\right)
T \int_{0}^{\infty} \dd f\, {\cal F}_D(f)\,,
\ee
where the spectral density of Fisher information is
\be\label{Fexplicit}
{\cal F}_D(f)=
3
\frac{\left[{\mathcal{R}}_3(f)
(1-n_I)\bar{I}(f)\right]^2}
{({\cal R}_A(f)\bar{I}(f)+N_A(f))
({\cal R}_T(f)\bar{I}(f)+N_T(f))}
\simeq
3
\frac{({\mathcal{R}}_3(f)(1-n_I))^2
\bar{I}(f)}
{{\cal R}_A(f)N_T(f)}\,,
\ee
where in the last step we assumed that the noise dominates the signal
in the $T$ channel but not in the $A$ channel.
To restore the dependence on the observing time and the cadence of the
observations, taking into account the discrete set of frequencies, one
should use the correspondence described for example in Eq.~(106) of
\cite{Pitrou:2024scp}, {that is $2 T \int_0^{f_{\rm max}} \dd f \leftrightarrow \sum_{n=-T f_{\rm max}}^{T f_{\rm max}}$ with $f_n = n/T$.}

\begin{figure}[ht!]
\centering
\includegraphics[width=0.6\textwidth]{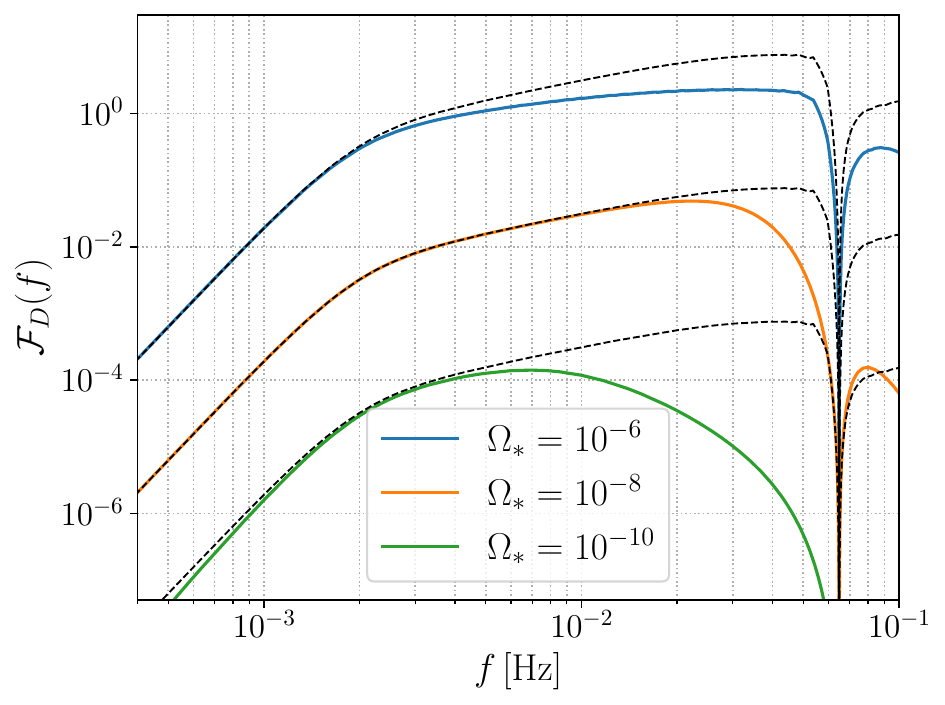}
\caption{\label{fig:Omegas2}
The function $\mathcal{F}_D(f)$ defined in Eq.~(\ref{Fexplicit}) for
several amplitude of the scale-free $h^2\Omega_{\rm GW}$ spectra (solid lines). The dashed
lines correspond to the approximation in the last equality of
Eq.~(\ref{Fexplicit}), where $\mathcal{F}_D$ becomes proportional to the
intensity.}
\end{figure}

In Fig.~\ref{fig:Omegas2} we show the function $\mathcal{F}_D(f)$ for a
vanilla inflationary spectrum. The function peaks around
$f\sim10^{-2}\,\mathrm{Hz}$ for sufficiently large signal amplitude.
This behaviour can be understood from Fig.~\ref{fig:response_functionA}:
the ratio of the dipole response function $\mathcal{R}_3$ to the noise
contribution is maximal near this frequency.

Physically, the singularity of the Fisher matrix in the static LISA case
reflects the fact that the detector cannot constrain a dipole oriented
normal to its plane. The sensitivity to a dipole arises from the
propagation time delay of gravitational waves between the interferometer
arms, which vanishes when the propagation direction is orthogonal to the
LISA plane.

Fortunately, the orientation of LISA is not fixed. Taking into account
its orbital motion, which induces a slow evolution of
$R^{\rm LISA}_{ij}(\tau_n)$, allows all dipole directions to be
constrained. This is discussed in detail in the next section.

\subsection{LISA in  motion}\label{LISAMoves}

 We now restore the time dependence of the orientation of LISA by taking
into account the rotation matrix $R^{\rm LISA}_{ij}(\tau_n)$. Since the
data from different time intervals $\tau_n$ are uncorrelated, the trace
over this index in the inner products reduces to a sum over $\tau_n$.
Compared to the static-LISA case, this effectively amounts to averaging
over the different orientations of LISA.  
The Fisher matrix then becomes
\be
F_{\beta_i \beta_j} \simeq
\langle \delta_{ij} - N^{\rm LISA}_i(\tau_n) N^{\rm LISA}_j(\tau_n)
\rangle_{\tau_n}
\,T \int_0^\infty \dd f \,{\cal F}_D(f)\,,
\ee
where the normal to the LISA plane in ecliptic
cooredinates evolves according to
\be \label{NormalLISA}
N^{\rm LISA}_i(\tau_n) = R^{\rm LISA}_{ij}(\tau_n) e_z^j\, .
\ee
Using the identity
\be\label{NLISAAv}
\langle N_{\rm LISA}^i N_{\rm LISA}^j \rangle_{\tau_n}
=
\frac{3}{8}\delta^{ij}
-
\frac{1}{8}e^i_z e^j_z\,,
\ee
which holds exactly when the mission duration is an integer number of
years, we obtain
\be\label{Fishervv}
{F}_{\beta_i \beta_j}
=
\left(
\frac{5}{8}\delta^{ij}
+
\frac{1}{8}e_z^i e_z^j
\right)
T \int_0^\infty \dd f\,{\cal F}_D(f)\,.
\ee
This matrix is now invertible, with inverse
\be
({F}^{-1})_{\beta_i \beta_j}
=
\left(
\frac{8}{5}\delta^{ij}
-
\frac{4}{15} e_z^i e_z^j
\right)
\left(
T \int_0^\infty \dd f\,{\cal F}_D(f)
\right)^{-1}.
\ee
The resulting estimator for the dipole velocity is therefore
\be
V^i =
({F}^{-1})_{\beta_i \beta_j} X^j\,,
\qquad
X^j \equiv
\sum_{\tau_n}
R^{\rm LISA}_{jk}(\tau_n)
\{ D^{k}; S_a S^\dagger_b \}_C\,.
\ee
Finally, the variance of the velocity amplitude estimator follows from
\be\label{sigmav}
\sigma_\beta^2 =
{\rm Tr}\!\left((F^{-1})_{\beta_i \beta_j}\right)
=
\frac{68}{15}
\left[
T \int_{0}^{+\infty} \dd f\,{\cal F}_D(f)
\right]^{-1}.
\ee
The Fisher-information spectral density ${\cal F}_D(f)$, together with
its large-noise approximation, is given in Eq.~\eqref{Fexplicit}.

\begin{figure}[h]
\centering
\includegraphics[width=0.48\textwidth]{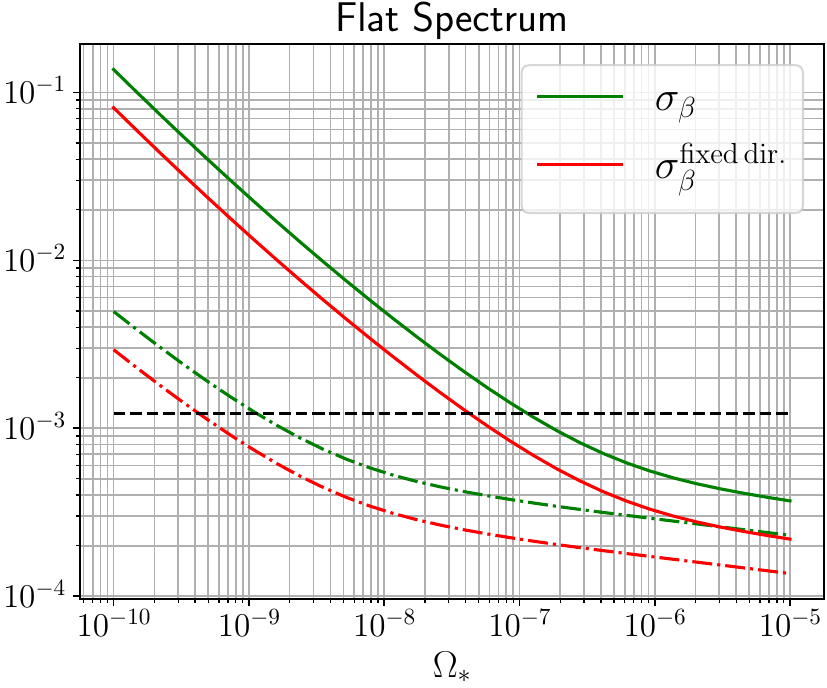}
\includegraphics[width=0.48\textwidth]{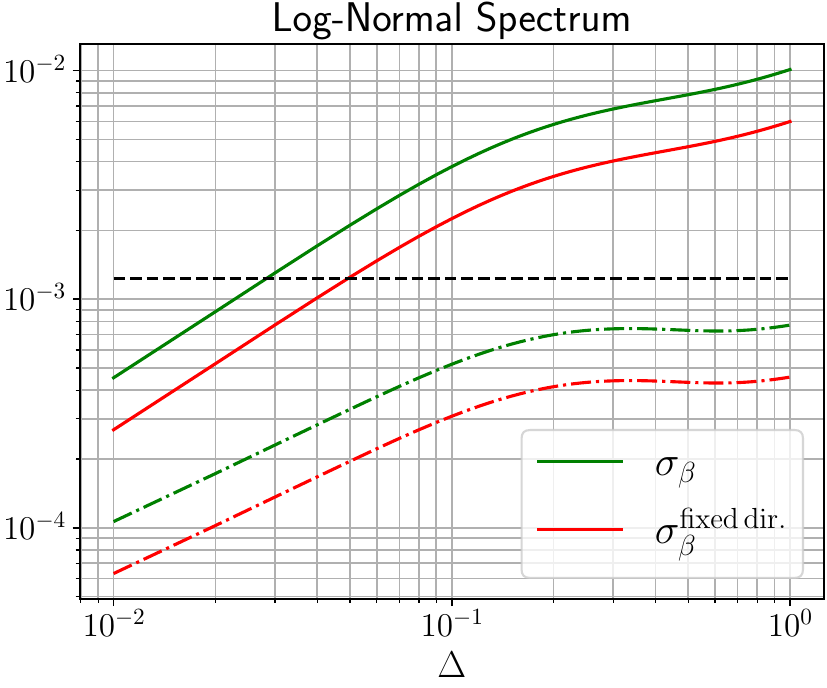}
\caption{Error on the modulus of the peculiar velocity (assuming/not assuming the direction fixed) as function of the amplitude $\Omega_*$ for a flat spectrum and as a function of the width $\Delta$ for a log-normal spectrum with area equal to $\mathcal{A}=10^{-8}$ and peaked at $f_{\rm p}=10\,\rm mHz$. We consider both the case in which we assume the velocity direction to be known and the case in which we let it vary. We checked that the two curves overlap once rescaled by the factor in Eq. \eqref{ImproveSigma}. On plain lines we use the \emph{fiducial} LISA model, while with dotted dashed we show the same curves obtained assuming the \emph{futuristic} LISA noise model. 
With  a blacked dotted line we plot the velocity estimated from CMB experiments $\beta=1.23\cdot  10^{-3}$.
}
\label{fig:scan_beta}
\end{figure}

\begin{figure}[h]
\centering
\includegraphics[width=0.49\textwidth]{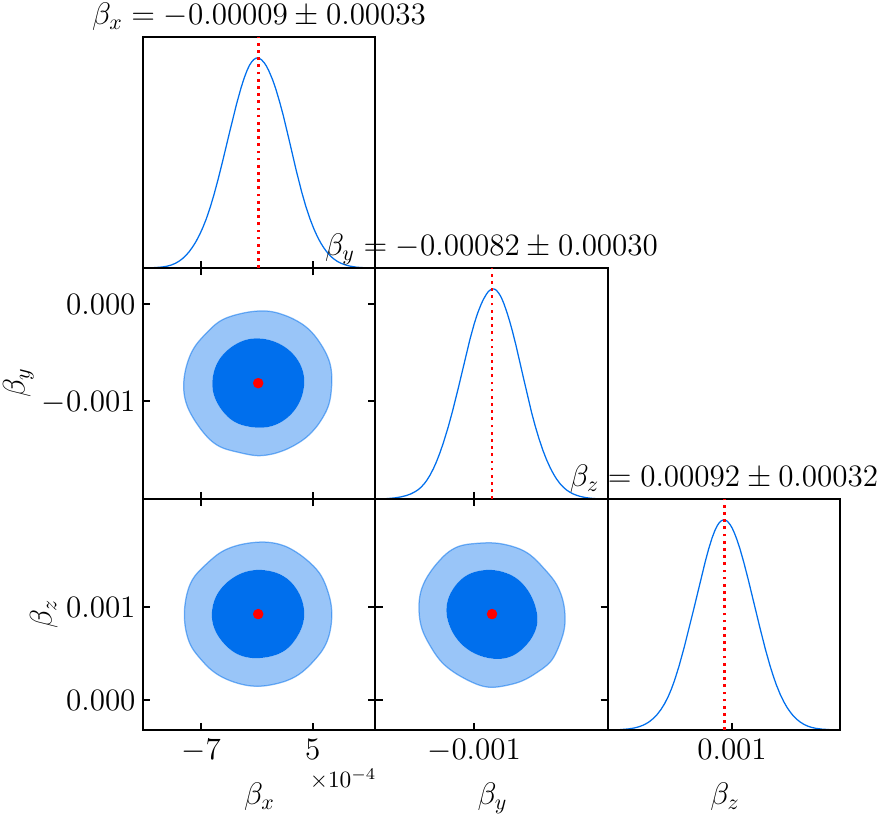}
\includegraphics[width=0.49\textwidth]{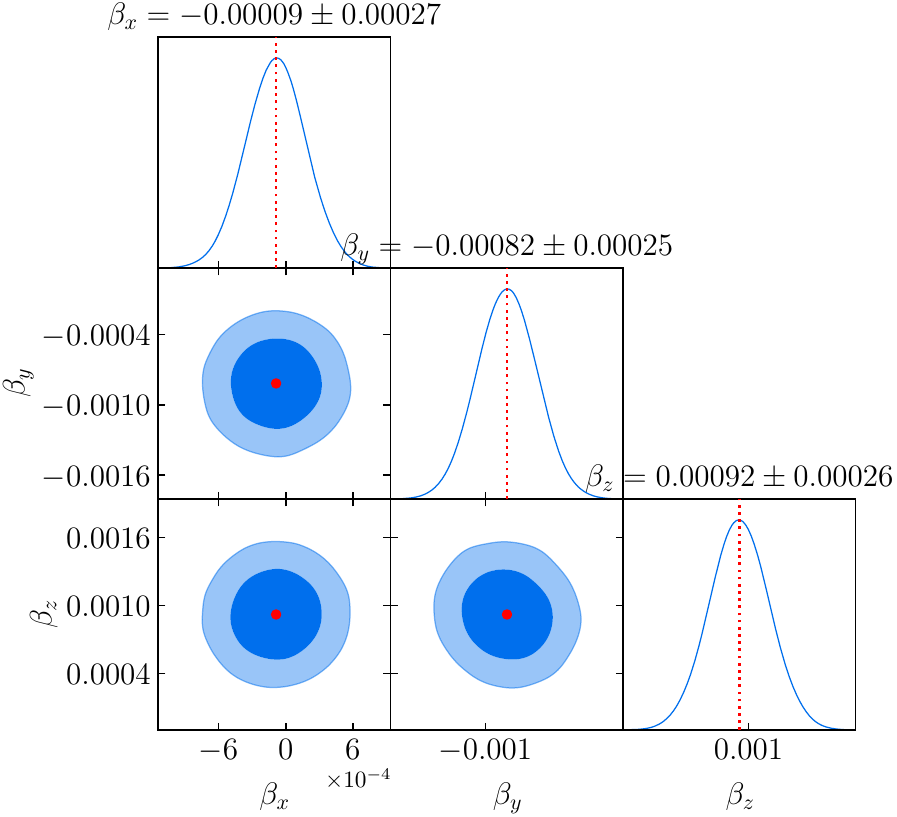}
\caption{Left: Corner plot for the components of the peculiar velocity in Galactic coordinates, assuming a flat spectrum with $\Omega_* =10^{-6}$. Right: Same plot for a primordial signal with log-normal template (given in Eq. \eqref{lognormal}) with parameters $f_p=10 \,\mathrm{mHz}$, $\Delta=10^{-2}$ and $\mathcal{A}=10^{-8}$. }
\label{fig:corner_beta_cartesian}
\end{figure}

\begin{figure}[h]
\centering
\includegraphics[width=0.48\textwidth]{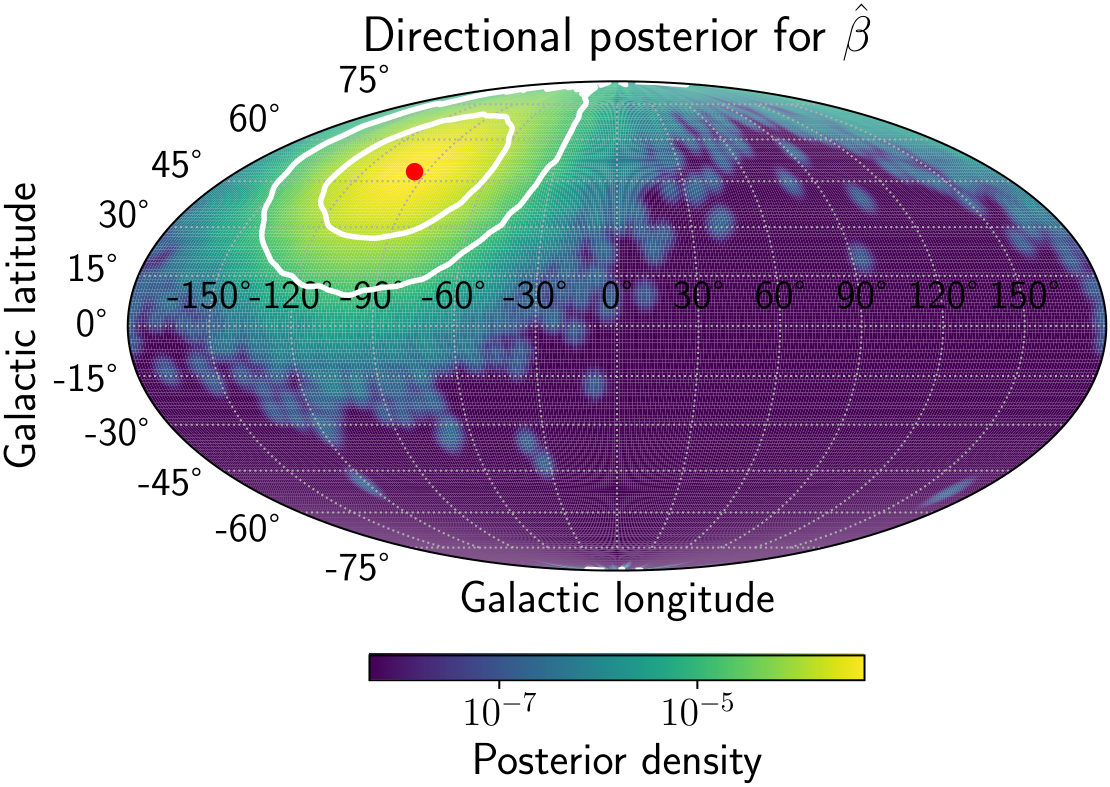} 
\includegraphics[width=0.48\textwidth]{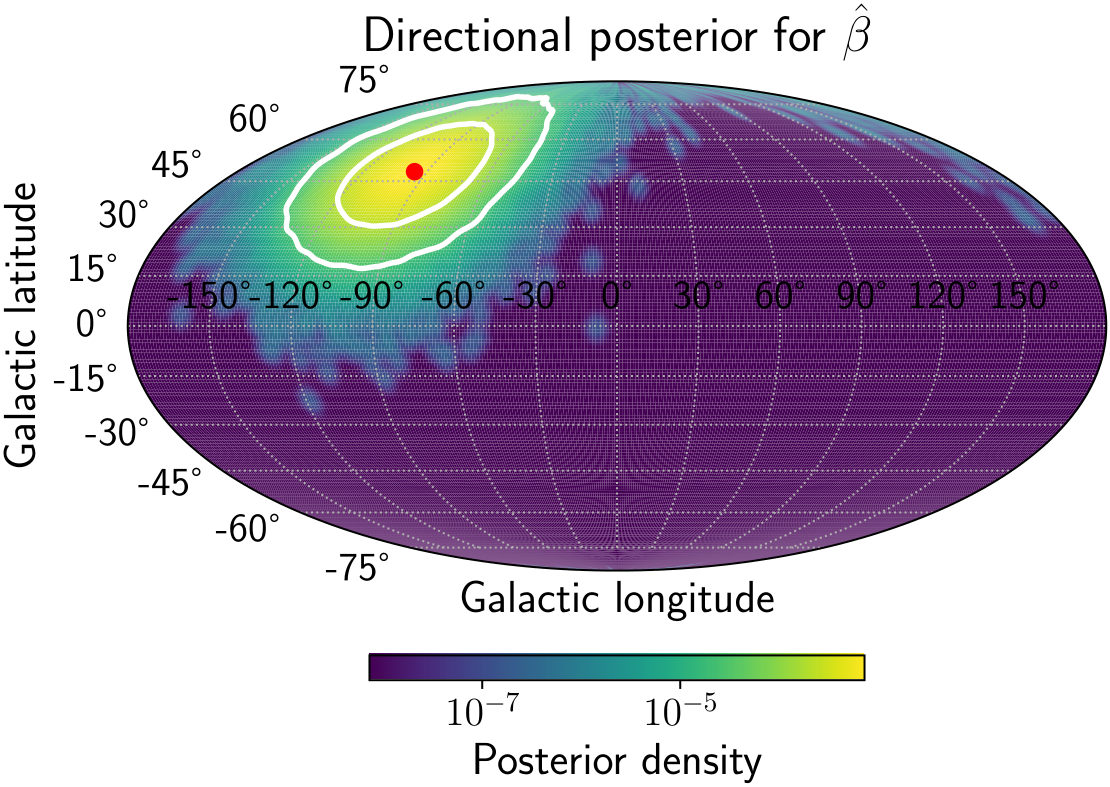}
\includegraphics[width=0.48\textwidth]{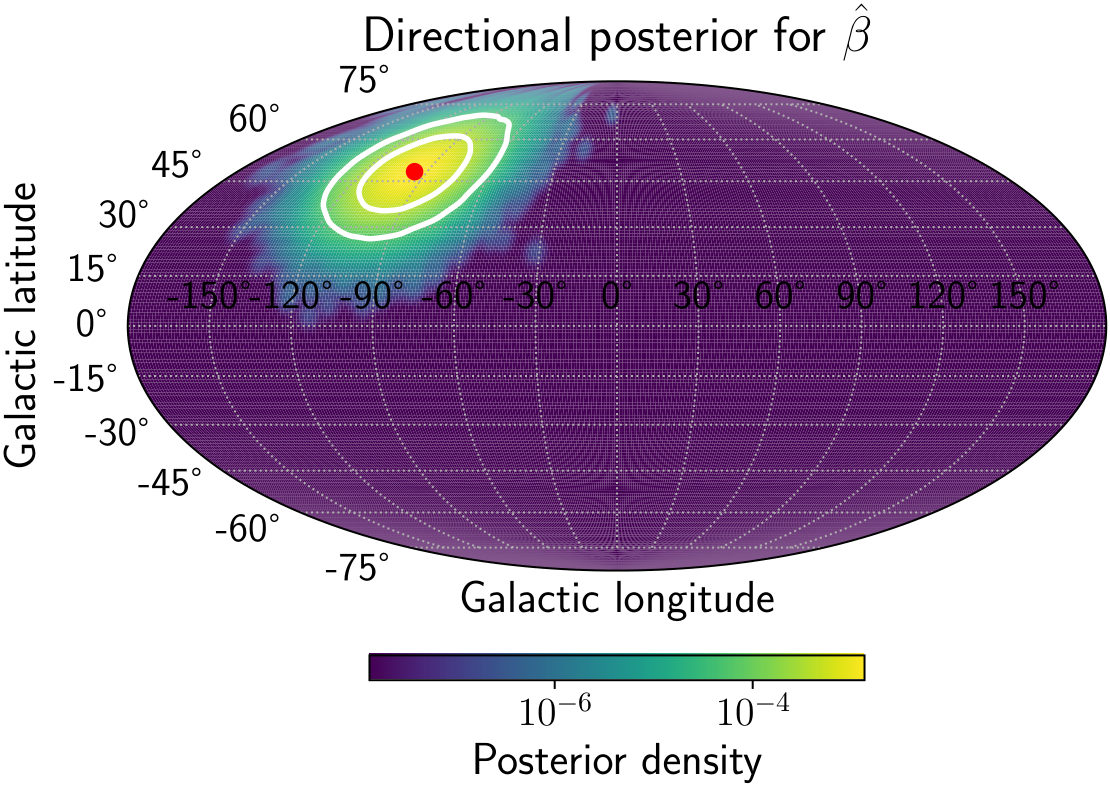}
\includegraphics[width=0.48\textwidth]{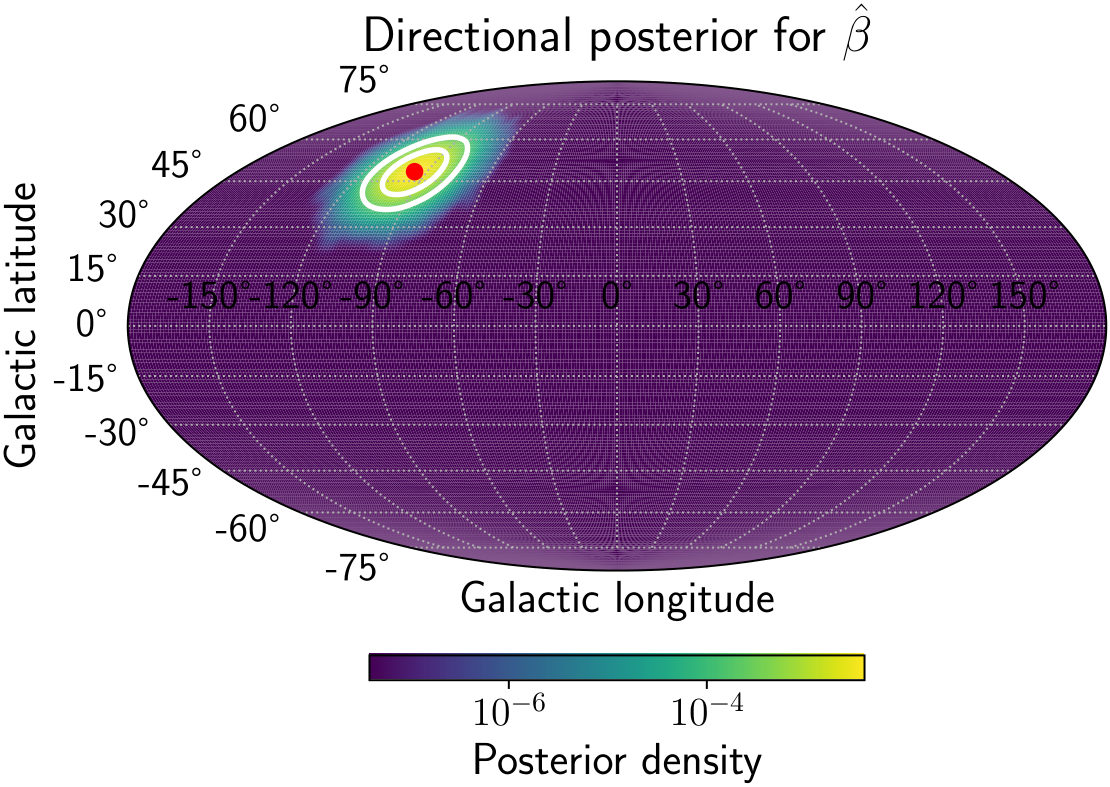}
\caption{Directional reconstruction of the peculiar-velocity unit vector $\hat{\beta}$ in Galactic coordinates for different benchmark primordial signals detailed below. The colour scale shows the directional probability density (per unit solid angle) on the sky obtained by propagating the Gaussian Fisher covariance of the Cartesian components $(\beta_x,\beta_y,\beta_z)$ to angular space, while the white contours indicate the 68\% (1$\sigma$) and 95\% (2$\sigma$) credible regions. The red marker denotes the CMB fiducial direction used in the Fisher evaluation. In all plots we use the \textit{fiducial} LISA noise model. Top left: Flat spectrum with $\Omega_\ast = 10^{-6}$, corresponding to the left corner plot in Fig.~\ref{fig:corner_beta_cartesian}.
Top right: 
Log-normal spectrum with same parameters as the right corner plot in Fig.~\ref{fig:corner_beta_cartesian}, i.e. $(f_{p},\Delta,\mathcal{A}) = (10\, \rm mHz, 10^{-2},10^{-8})$. 
Bottom left: Log-normal spectrum with parameters $(f_{p},\Delta,\mathcal{A}) = (10\, \rm mHz, 5\times 10^{-2},10^{-6})$ (same amplitude as in the top-left panel, but with a narrower spectrum). Bottom right: Log-normal spectrum with parameters $(f_{p},\Delta,\mathcal{A}) = (10\, \rm mHz, 10^{-2},10^{-7})$ (same width as in the top-right panel, but with a higher amplitude).
}
\label{fig:corner_beta_polar}
\end{figure}

If the dipole orientation is assumed to be known, the Fisher matrix can be restricted to that direction.
 Denoting by $\hat \beta^i$ the unit
vector along the dipole direction, the relevant Fisher information is
\be\label{defFfixed}
{F}^{\rm fixed} = {F}_{\beta_i \beta_j} \hat{\beta}^i \hat{\beta}^j\,.
\ee
The corresponding variance of the velocity estimator becomes
\be\label{ImproveSigma}
\left.\sigma_\beta^2\right|_{\text{fixed direction}}
=
\frac{1}{F^{\rm fixed}}
=
\sigma_\beta^2 \times
\frac{6}{17}
\left(
1+\frac{1}{5}\cos^2\theta_{\rm dip}
\right)^{-1},
\ee
where $\theta_{\rm dip}$ denotes the angle between the dipole direction
and the normal to the ecliptic plane. As expected, this variance is
smaller than the one obtained when the dipole direction is left
completely free, corresponding to an improvement by a factor of order
$\sim 0.3$.

In Fig.~\ref{fig:scan_beta} we show the variance of the reconstructed
velocity as a function of the background amplitude for two examples: a
scale-free primordial spectrum with amplitude 
$\Omega_\star$ 
and a log-normal spectrum with GW density 
\be\label{lognormal}
h^2\Omega_{\text{GW}}\,=\,\frac{\mathcal{A}}{\sqrt{2\pi\Delta^2}}\exp{\left[-\frac{\ln^2(f/f_p)}{2 \Delta^2} \right]}\,.
\ee
We assume
a dipole amplitude equal to that measured in the CMB,
$\beta/c = 1.23\times10^{-3}$.

For a flat spectrum, and assuming that the direction of the motion is
known and fixed to the value measured by CMB experiments, we find that
detectability can be achieved for\footnote{In deriving these detectability thresholds, we neglect the galactic foreground. This approximation is expected to be adequate for the large background amplitudes relevant for velocity reconstruction, for which the extra-galactic or primordial contribution should dominate over the galactic foreground in the LISA band. A more refined treatment of foreground contamination could affect the precise numerical value of the threshold, but is not expected to change the qualitative conclusion. This is particularly true for extra-galactic signals extending beyond $1\,\mathrm{mHz}$, such as those considered here, where the galactic foreground contamination is expected to decrease.
} $\Omega_* \gtrsim 5\times 10^{-8}$ with
\emph{fiducial} LISA, and for $\Omega_* \gtrsim 5\times 10^{-10}$ with the
\textit{futuristic} instrument. These results for the \emph{fiducial} LISA are in full agreement with the
simulations of \cite{Heisenberg:2024var}.\footnote{In particular, fixing $H_0\simeq 67\, \mathrm{km\,s^{-1}\,Mpc^{-1}}$, we also recover the threshold $\Omega_{\rm GW} \gtrsim 5\times 10^{-8}$ quoted in that work by adopting what the authors label as the best-estimate noise (namely, a more optimistic noise level than
\cite{Flauger:2020qyi, Babak:2021mhe, Robson:2018ifk}). This corresponds to lowering the OMS parameter $P$ in Eq.~\eqref{noise} from $15$ to $6.35$, following \cite{Bayle:2022okx}.} 

They are also consistent with
Fig.~10 of \cite{Bartolo:2019oiq}, although we note an apparent
inconsistency with the detection threshold quoted in the abstract of
that work. 
Finally, one can apply an argument similar to the one presented after Eq.~\eqref{DefSNR2Omega2} to compute the SNR associated to the dipole amplitude. It is defined by 
\be\label{SNRdipole}
\text{SNR}^2_\beta \equiv \frac{\beta^2}{\sigma_\beta^2}= \beta^2 T \frac{15}{68} \int_0^{\infty} \dd f\, \mathcal{F}_D\,.
\ee
Using that in the limit of signal domination $\mathcal{F}_D\sim 3$, therefore $T\int_0^{\infty} \dd f\, \mathcal{F} \simeq 3 T f_{\rm max} \simeq 3.7\times 10^7$ and we conclude that we cannot hope to detect a dipole velocity smaller than $\simeq 3 \times 10^{-4}$. We find indeed in Fig.~\ref{fig:scan_beta} (left panel) that there is a lower bound on $\sigma_\beta$ with this order of magnitude. As long as the spectrum is featureless, the only way to improve this bound is by increasing the LISA operating time $T$. In contrast, a narrow spectrum can evade this lower limit (see the right panel of Fig.~\ref{fig:scan_beta}) owing to the contribution of the spectral index in the numerator of Eq.~\eqref{Fexplicit}.

In Fig.~\ref{fig:corner_beta_cartesian} we present the Fisher constraints on the Cartesian velocity components $(\beta_x,\beta_y,\beta_z)$ in the galactic frame for two different signal models.\footnote{Notice that our results are in agreement with the requirement that the spectral index should not be larger than $\beta^{-1}$, which is necessary for the  analysis to be consistent
at leading order in $\beta$ \cite{Mentasti:2025ywl}.} In Fig.~\ref{fig:corner_beta_polar}, we show the directional reconstruction of the peculiar-velocity unit vector in Galactic coordinates. The figure makes it visually clear that the reconstruction improves as the spectrum becomes narrower. In particular, the top-left panel shows the reconstruction for a flat spectrum with $\Omega_*=10^{-6}$, while the bottom-left panel displays the corresponding case for a log-normal spectrum with area $\mathcal{A}=10^{-6}$ and width $\Delta=5\times10^{-2}$. Along similar lines, the top-right panel corresponds to a log-normal spectrum with $\mathcal{A}=10^{-8}$ and $\Delta=10^{-2}$, whereas the bottom-right panel shows a log-normal spectrum with the same width, $\Delta=10^{-2}$, but a larger amplitude, $\mathcal{A}=10^{-7}$.

\section{Kinematic dipole as degeneracy breaker}\label{deg_breaker}

\subsection{Heuristic argument}

As already mentioned, the detection of extragalactic stochastic background components with LISA presents two major challenges. First, in the LISA frequency band one expects a strong galactic foreground that may obscure signals of cosmological interest \cite{Boileau:2025jkv,Staelens:2023xjn,Hofman:2024xar,Criswell:2024hfn}. 
Second, uncertainties in the modelling of instrumental noise further complicate the analysis (see e.g. \cite{Adams:2010vc,Muratore:2022nbh,PhysRevD.109.042001,Baghi:2023qnq,Alvey:2024uoc}). These effects can lead to significant degeneracies between primordial, galactic, and noise contributions.

Motivated by this, we explore the following idea: since cosmological and extragalactic signals exhibit a kinematic modulation, this feature can be used to break degeneracies with galactic foregrounds and instrumental noise, which are not expected to share the same anisotropic signature.

To illustrate this mechanism, we consider a simplified setup. 
We assume the presence of two components: a primordial (or any extragalactic) signal that exhibits a kinematic dipole, and an additional, degenerate component, which does not exhibit such modulation. For the purposes of this argument, we further assume that the component without a dipole has the same spectral shape as the primordial signal, thereby maximising the degeneracy at the monopole level.
Within this setup, we adopt the following simplified notation for the total covariance in Eq.~\eqref{CgeneralXX}:
\be
C = M + D + N = M_p + M_g + D + N\,,
\ee
where all channel indices and frequency dependences are left implicit, and $M_p$ and $M_g$ denote, respectively, the monopole contribution of the primordial/extragalactic signal and that of an additional component, interpretable as either a galactic foreground or an unknown noise contribution.
For simplicity, all quantities are understood to be expressed in Fourier space. We further assume that the only parameters to be constrained are the overall amplitude of the primordial component, $\Omega_*$, and the amplitude of the galactic foreground, $\Omega_{\mathrm{fg}}$.
As in the previous section, we approximate $C$ with $M+N$ in the matrix defining the Fisher inner products. It can be shown that if $A=\ii{\cal A}$ with ${\cal A}$ antisymmetric,
and if $B$ is real symmetric (so that both matrices remain Hermitian),
then $\{A;B\}_C=0$ for any positive-definite symmetric $C$.  This will allow us to simplify the following treatment since it implies schematically \be\label{magicproperty}
\{\partial M; \partial D\}_{M+N}=0.
\ee 
In order to alleviate the notation, we will omit subscripts in the inner products and define
\be
\bar{s} \equiv \frac{\partial M_p}{\partial\log\Omega_*} ;\qquad d \equiv \frac{\partial D}{\partial\log\Omega_*} ;\qquad g\equiv \frac{\partial M_g}{\partial \log\Omega_{\mathrm{fg}}};\qquad s \equiv \bar{s}+d. 
\ee
The $2\times 2$ Fisher matrix can be written as 
\be
{F}_{\Omega \Omega}= \{s;s\}=\{\bar{s};\bar{s}\}+\{d;d\}, \qquad {F}_{gg}= \{ g;g \},\qquad {F}_{\Omega g}=  \{ s ;g \} =\{ \bar{s} ;g \}.
\ee
In this simplified parameter space, the marginalized information on $\Omega_*$ sometimes referred to as the Schur complement is (in absence of priors)
\be
 {F}^{\text{marg}}_{\Omega \Omega} \equiv \frac{1}{\sigma^2_{\log\Omega_*}}= \frac{1}{(F^{-1})_{\Omega\Omega}}={F}_{\Omega \Omega}(1-\rho^2)\,,\quad \text{where}\quad\rho\equiv \frac{{F}_{\Omega g}}{\sqrt{{F}_{\Omega \Omega}{F}_{gg}}}\,.
\ee
If the two monopole components are proportional, i.e. $M_g = \kappa M_p$, where $\kappa$ is a constant, then in the absence of the kinematic dipole we have $F_{gg} = \kappa^2 F_{\Omega\Omega}$ and $F_{g\Omega} = \kappa F_{\Omega\Omega}$. It follows that $\rho = 1$ and $F^{\text{marg}}_{\Omega\Omega} = 0$, corresponding to a perfect degeneracy.
Realistically, observational and astrophysical constraints can be used to reduce the allowed parameter space of the foreground component, which amounts to adding a prior to the Fisher matrix. Introducing a Gaussian prior on the foreground amplitude, with associated precision $\sigma_{\text{p}}^{-2}$, shifts the $F_{gg}$ Fisher element according to $F_{gg} \to F_{gg} + \sigma_{\text{p}}^{-2}$. 
Then the marginalised variance will be finite and tend to the prior.

When the dipole is included in the analysis, it also acts as a degeneracy breaker, preventing the marginalised Fisher information from vanishing even in the absence of (or in the presence of weak) priors. To illustrate this, it is convenient to decompose the primordial contributions using a projector on the foreground component $g$:
\be
x= (1-\Pi_g)x + \Pi_g x\equiv x_{\perp} + x_{||}\,,\qquad \mathrm{with}\qquad \Pi_g\, x \equiv \frac{\{g;x \}}{\{g,g \}} g\,.
\ee
Through simple manipulations one can write in general
\be
\,F^{\rm marg}_{\Omega\Omega}\equiv\frac{1}{\sigma^2_{\log\Omega_*}} = \,\{ s_{\perp}; s_{\perp} \} +\frac{1}{1+\{ g; g\}\sigma^2_p} \{ s_{||};s_{||}\} = \{ s; s \}  (1-\rho^2)\,,
\ee
where
\be
\rho^2 \equiv \frac{\{ s_{||}; s_{||} \} } {\{ s; s  \}}\frac{\{g;g\}}{\sigma^{-2}_p+ \{g;g\}}\,,
\ee
from which a clear geometrical interpretation emerges. Several limiting cases are particularly instructive  which we will illustrate in the following.

If the two monopole components are proportional and the dipole is absent, so that $s=\bar{s}=s_{||}$, (perfect degeneracy). In this case, the marginalised Fisher information is entirely controlled by the priors:
\be
F^{\rm marg}_{\Omega\Omega} = \frac{1}{1+F_{gg}\sigma_p^2}\,F_{\Omega\Omega}
\simeq \frac{F_{\Omega\Omega}}{\sigma_p^2 F_{gg}}
= (\kappa \sigma_{\mathrm{p}})^{-2},
\ee
where in the second-to-last step we expanded for small $\sigma_{\mathrm{p}}^{-2}$, corresponding to wide priors. If no priors are imposed, $\sigma_p^{-2}=0$, and the primordial component has a kinematic dipole whereas the galactic component does not, so that $d=d_{\perp}$, while the monopole sector remains perfectly degenerate, $\bar{s}=\bar{s}_{||}$, then
\be
F^{\rm marg}_{\Omega\Omega} = \{ d; d\}\,.
\ee
If priors are included and the extragalactic/primordial component has a dipole whereas the galactic one does not, so that $d=d_{\perp}$, while the monopole sector is still perfectly degenerate, $\bar{s}=\bar{s}_{||}=s_{||}$, then
\be
F^{\rm marg}_{\Omega\Omega} = \{ d; d\} + \frac{1}{1+F_{gg}\sigma_p^{-2}} \{ \bar{s};\bar{s}\}\,.
\ee
Note that in this case $F_{\Omega\Omega}\equiv\{ s;s\}\neq \{\bar{s};\bar{s}\}$.
In general, it is the component of the dipole orthogonal to the foreground space that breaks the degeneracy, increases the marginalised Fisher information, and therefore improves parameter reconstruction. The same applies to non-standard noise templates: if a noise feature mimics a signal, the same degeneracy issue arises. Including a dipole component helps breaking these degeneracies in the same way as illustrated above. In the next section, we will explicitly demonstrate this with concrete examples. 
The following analysis should be regarded as a proof of concept rather than as a fully realistic end-to-end forecast. The simplified foreground and noise models considered here are intended to illustrate the basic mechanism: if a degeneracy is present at the level of the monopole between a primordial background and a component without a corresponding kinematic dipole, the dipole provides additional information that can help break this degeneracy. A quantitative assessment of this effect in the presence of realistic foregrounds, intrinsic anisotropies, and imperfect subtraction is left for future work.

\subsection{Degeneracy with a foreground component}

We now place our idea on a more solid footing. We assume that our goal is to simultaneously constrain the parameters of the galactic and extragalactic background components, and we demonstrate how the inclusion of a dipole term helps to improve constraints on the model parameters. We keep the dipole velocity fixed, assuming it to be the value measured by CMB experiments. Fisher-matrix elements for all sub-blocks are derived and presented in Appendix~\ref{FisherDet}. The only non-trivial contribution comes from the terms involving the dipole. These always appear in a coupled form, thanks to the property discussed in Eq.~\eqref{magicproperty}. Denoting by $\mu$ and $\nu$ two generic parameters, we find, recalling the definition in Eq.~\eqref{DefDXXi}, 
\begin{align}\nonumber
F_{\mu\nu}\supset \{\partial_{\mu} D,\partial_{\nu}D\}&\propto \left(\sum_{O=A,E}\langle D^{i}_{OT}D^{j}_{TO}R_{mi}^{\rm LISA}(\tau_n)R_{nj}^{\rm LISA}(\tau_n) \rangle_{\tau_n}\right)\\\nonumber &\qquad\times\partial_{\mu}[\beta^m (1-n_I)\bar{I}]\partial_{\nu}[\beta^n (1-n_I)\bar{I}]\\[1mm]
&=3 (\mathcal{R}_3(f))^2\left(\frac{5}{8}\delta_{mn}+\frac{1}{8}e^{m}_{z} e_z^n\right)\partial_{\mu}[\beta^m (1-n_I)\bar{I}]\partial_{\nu}[\beta^n (1-n_I)\bar{I}],
\end{align}
where, as before, $\langle...\rangle_{\tau_n}$ denotes the average over time segments, and in going from the first to the second line we have used Eqs.~\eqref{sumAO}, \eqref{NormalLISA}, and \eqref{NLISAAv}. If $\mu$ and $\nu$ are components of the peculiar velocity $\boldsymbol{\beta}$, we recover the result of the previous section, Eq.~\eqref{Fishervv}. If instead they are signal parameters, the above expression reduces to
\be
F_{\mu\nu}\supset \frac{3 (\mathcal{R}_3(f))^2}{8}(5\beta^2 +\beta_z^2)\partial_{\mu}[(1-n_I)\bar{I}]\partial_{\nu}[(1-n_I)\bar{I}],
\ee
where we recall that $\beta_z$ is the $z$ component of the peculiar velocity in the ecliptic frame.
\begin{figure}[h]
\centering
\includegraphics[width=0.52\textwidth]{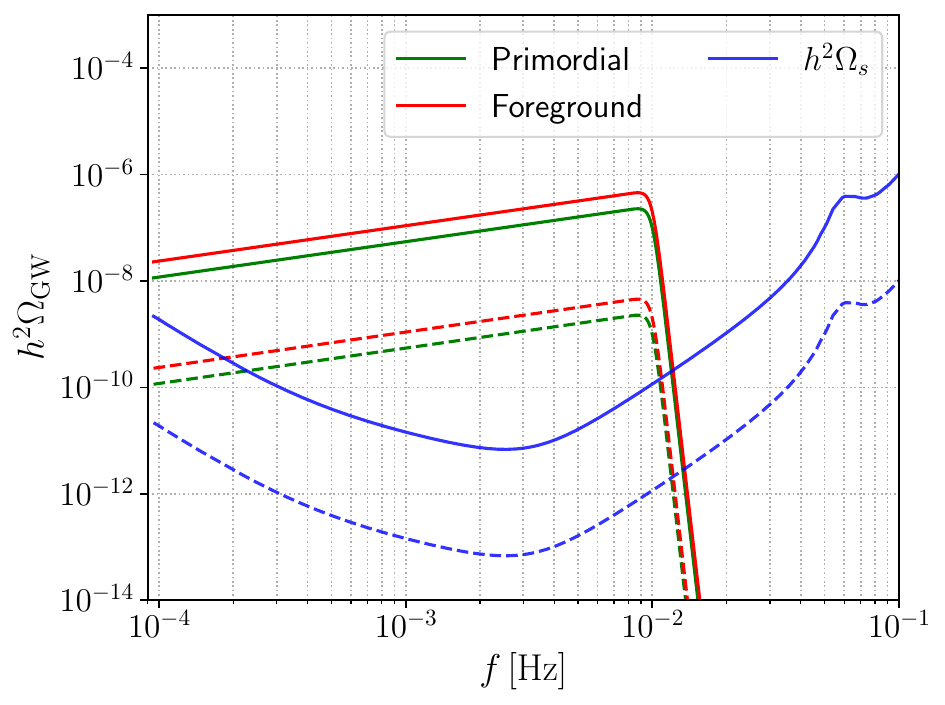}\quad
\includegraphics[width=0.45\textwidth]{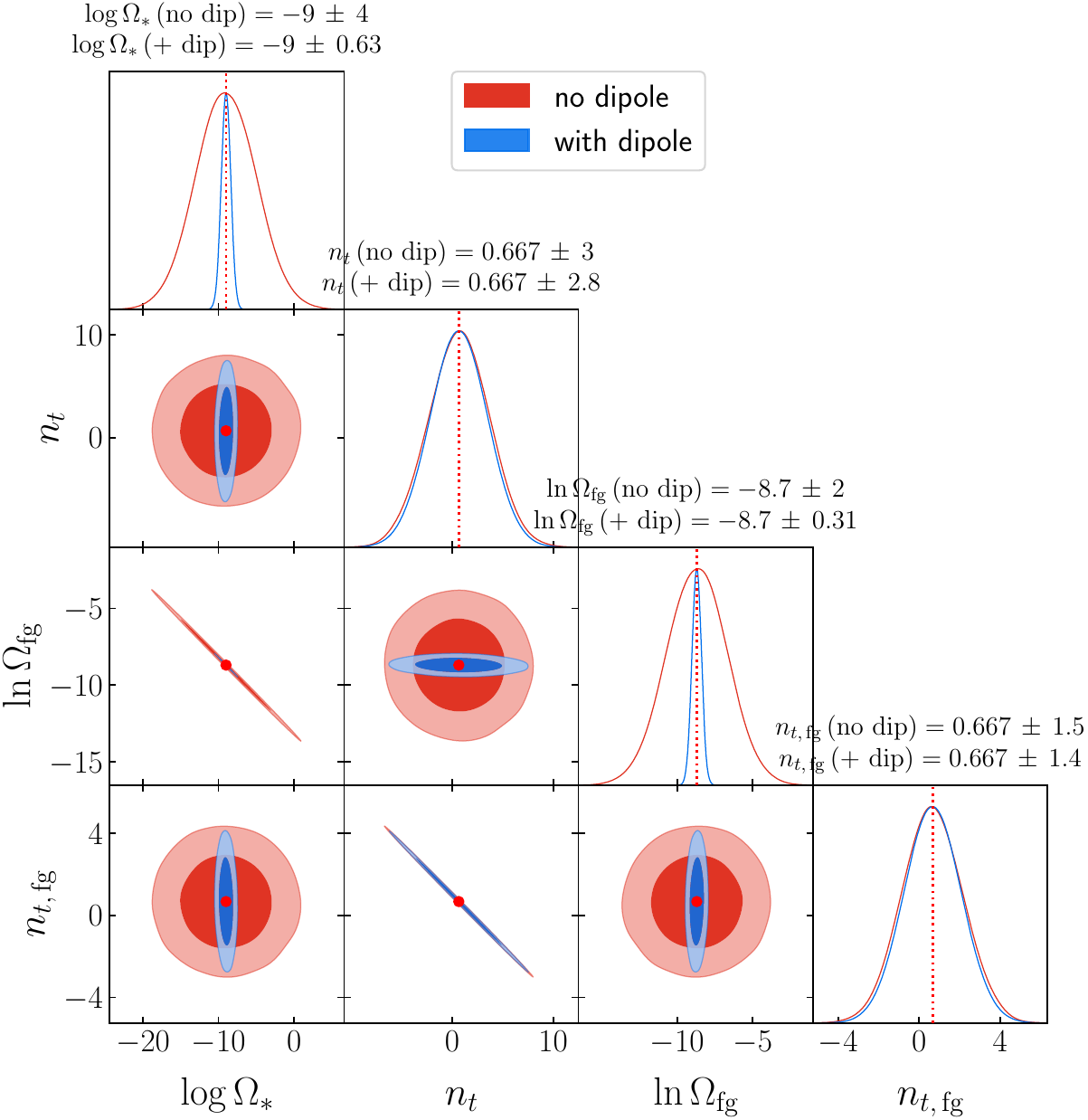}
\caption{Overlapping signals. Left: Solid lines denote the primordial monopole component of a signal with peak amplitude $\Omega_* = 10^{-7}$, superimposed on a galactic foreground-like component with the same shape and amplitude $\Omega_{\rm fg} = 2 \times 10^{-7}$, and shown together with the sensitivity corresponding to the \textit{fiducial} LISA noise in Eq.~(\ref{NaiveOmegas}). Dashed lines denote the same primordial and foreground components, both reduced by two orders of magnitude, along with the sensitivity estimates for the \textit{futuristic} LISA configuration described in the main text. Right: Corner plot for the primordial and foreground signals (with and without kinematic dipole) for a realistic foreground level and in the \textit{futuristic} LISA scenario (dashed line curves on the left panel). The error estimates is equivalent to the solid lines scenario of the left panel except for the different (rescaled) fiducial values.}
\label{degeneracy_plot_signals}
\end{figure}

\begin{figure}[h]
\centering
\includegraphics[width=0.47\textwidth]{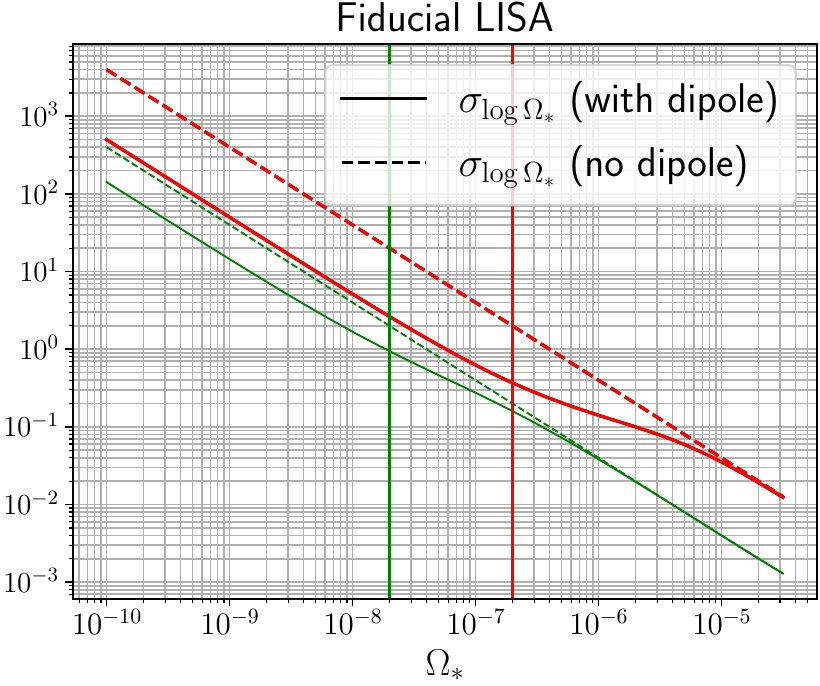}\quad
\includegraphics[width=0.47\textwidth]{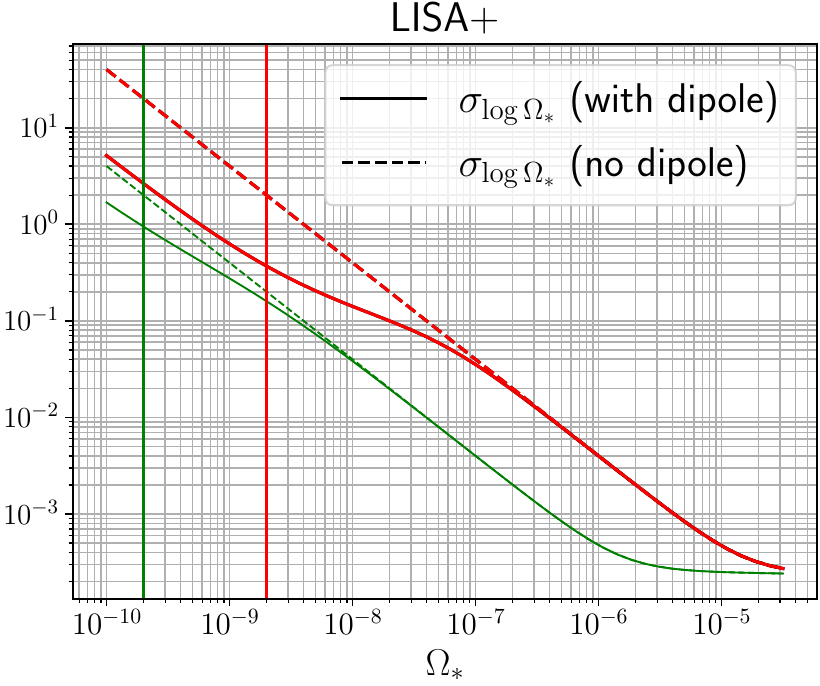}
\caption{
Marginalised variance of the logarithmic amplitude of the primordial signal, in the presence of degeneracy with a foreground component (see text), as a function of the fiducial primordial amplitude, shown with and without the dipole contribution. The fiducial spectral indices of both the primordial and galactic components are fixed to $2/3$. Vertical lines indicate the adopted fiducial foreground amplitudes, with colors matching the corresponding marginalised-variance curves. Left: \textit{fiducial} LISA noise, with $\Omega_{\rm fg}=2\times10^{-7}$ (red; as in Fig.~\eqref{degeneracy_plot_signals}) and $\Omega_{\rm fg}=2\times10^{-8}$ (green). Right: \textit{futuristic} noise setup, with $\Omega_{\rm fg}=2\times10^{-9}$ (red) and $\Omega_{\rm fg}=2\times10^{-10}$ (green). As explained in the text, in the low-amplitude regime, $\Omega_*<\Omega_{\rm fg}$, the separation between the curves with and without the dipole depends only on the ratio of the noise amplitude to the foreground amplitude. Because this ratio is kept fixed between the left and right panels, the ratio between the corresponding $\sigma_{\log \Omega_*}$ curves is unchanged in the low-$\Omega_*$ regime.}
\label{fig:degeneracy_scans1}
\end{figure}

We consider benchmark cases in which the primordial signal and the galactic foreground components have similar spectral shapes, both being described by a power law 
\be
h^2\Omega_{\mathrm{GW}}=\Omega_{i}\left(\frac{f}{f_*}\right)^{n_i},
\ee
with the same sharp cutoff, and with parameters labeled as $\{\Omega_*,n_t\}$ and $\{\Omega_{\mathrm{fg}},n_{t,\mathrm{fg}}\}$ respectively. We also take $f_*=0.01\,\mathrm{Hz}$ fixed for simplicity.
We assign a prior $\sigma_p = 2$ to the logarithm of the foreground amplitude and a prior $\sigma_p = 1.5$ to the foreground spectral index.  As discussed previously, priors break cases of perfect degeneracy even before the dipole contribution is included. 
Two benchmark signals of this type are shown in Fig.~\ref{degeneracy_plot_signals} (right). 
In both cases, the spectral indices of the two components are identical $n_t=n_{t,\mathrm{fg}}=2/3$. 
In solid lines, we choose amplitudes $\{\Omega_*,\Omega_{\rm fg}\}=\{10^{-7},2\times 10^{-7}\}$ and assume the \textit{fiducial} LISA noise model.
This situation corresponds to an unrealistically large foreground covering a primordial signal. Nevertheless, it is instructive to consider this case in view of a possible mission following LISA, since the parameter-estimation results are unchanged if all components (signal, foreground and the instrumental noise) are rescaled by the same factor.  We consider the \emph{futuristic} LISA scenario (the two noise parameter $P$ and $A$ reduced by a factor $10$, keeping the mission time of the nominal LISA experiment).\footnote{Obtaining such a noise reduction is quite unrealistic for the LISA nominal mission: a more realistic picture would give a noise reduction of a factor $\sim 20-30$, hence one would have to increase the mission time of a factor $\sim 4$ to get the same picture as the one described here.} 
This situation is depicted with a dashed lines in Fig.~\ref{degeneracy_plot_signals}, where we consider the same primordial signal and foreground shapes, but a more realistic foreground amplitude, i.e. $\{\Omega_*,\Omega_{\rm fg}\}=\{10^{-9},2\times 10^{-9}\}$. 
The corresponding uncertainties, shown in the right panel of Fig.~\ref{degeneracy_plot_signals}, are identical to those obtained for the solid-line configuration (primordial signal, foreground, and noise) shown in the left panel of Fig.~\ref{degeneracy_plot_signals}. In this case, the inclusion of the dipole constraint clearly helps breaking the degeneracy in parameter space. Since the dipole depends only on the spectral index of the cosmological background, including it improves the reconstruction of the shape of the subdominant cosmological signal, which in turn breaks the degeneracy between the two spectral indices and significantly improves the determination of the foreground slope. This benchmark case can be viewed as a representative point in the two-dimensional parameter scans that we now describe.
In Fig.~\ref{fig:degeneracy_scans1} on the left panel, we present the marginalised variance of the primordial amplitude as a function of its fiducial value, with and without including the dipole in the analysis, for two different values of the foreground amplitude and where the fiducial spectral indexes are taken to be identical $n_t=n_{t,\,\mathrm{fg}}=2/3$. We assume the \emph{fiducial} LISA noise model on the left panel and the \emph{futuristic} LISA on the right. 

We observe two main features once the dipole is included. First, the dipole improves the reconstruction when the fiducial primordial signal is smaller than, or up to roughly one order of magnitude larger than, the foreground amplitude. As expected, at sufficiently large primordial amplitude $\Omega_\ast$, the diagonals/monopole contribution to the Fisher are signal dominated, the Fisher information on $\log\Omega_\ast$ becomes large, and the amplitude is already well reconstructed from monopole terms alone. Then the ``with dipole'' and ``no dipole'' error curves merge. Second, focusing on the region where $\Omega_* \leq \Omega_{\mathrm{fg}}$, the relative improvement depends on the ratio between the foreground amplitude and the noise level. As expected, for fixed noise, the dipole provides a larger improvement when the foreground is stronger, since in that case the monopole reconstruction is more severely degraded.

The situation is therefore qualitatively identical in the \emph{futuristic} LISA scenario shown in the right panel of Fig.~\ref{fig:degeneracy_scans1}, with foreground levels lowered by two orders of magnitude relative to the left panel (recall that the $A$ and $P$ noise parameters are lowered by a factor of $10$ and they enter quadratically in the noise), except that the horizontal axis is shifted to the right by a factor of $100$.

\begin{figure}[h]
\centering
\includegraphics[width=0.47\textwidth]{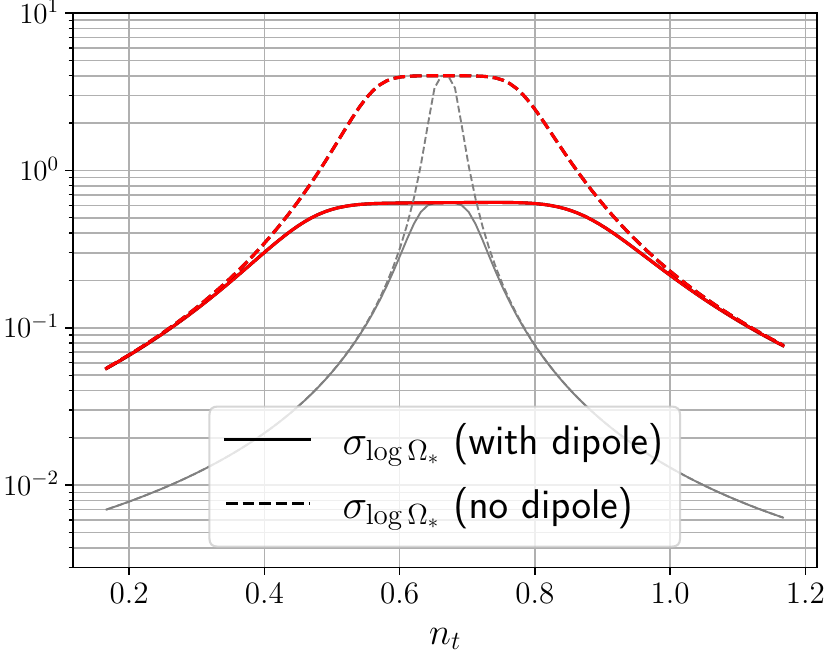}
\includegraphics[width=0.47\textwidth]{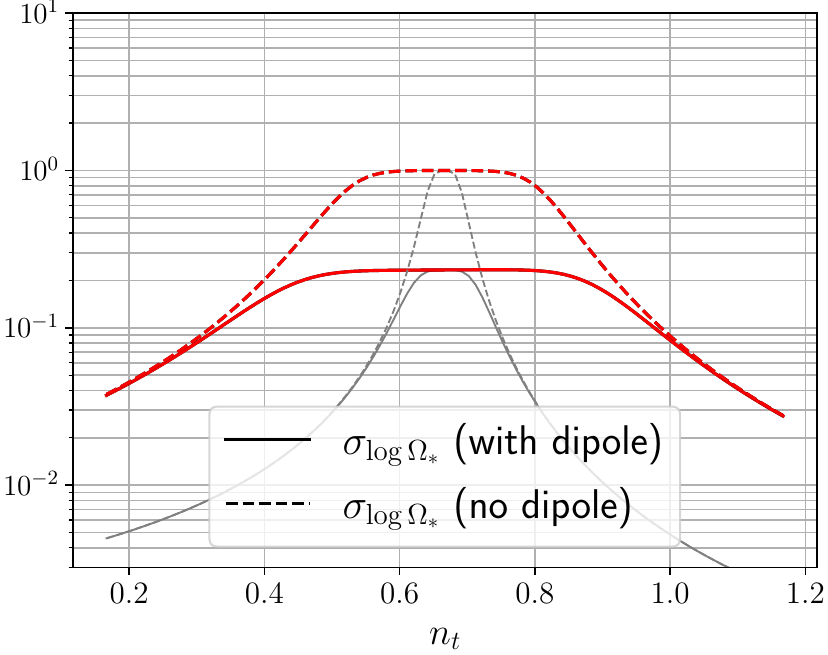}
\caption{Marginalised variance of the logarithmic amplitude of the primordial signal, in the presence of degeneracy with a foreground component (see text), as a function of its spectral index. The red and grey curves correspond to wide priors ($\sigma_p = 1.5$) and very narrow priors ($\sigma_p \ll 1$), respectively, on the foreground spectral index $n_{t,\rm fg}$, whose fiducial value is set to $2/3$. In both panels, the fiducial foreground amplitude is fixed to $\Omega_{\rm fg} = 3 \times 10^{-7}$ and the \textit{fiducial} noise model is assumed. The fiducial primordial amplitude is set to $\Omega_* = \Omega_{\rm fg}/2$ in the left panel and to $\Omega_* = 2\,\Omega_{\rm fg}$ in the right panel. The plots would be identical for a realistic foreground amplitude, $\Omega_{\rm fg} \simeq 10^{-9}$, and the \textit{futuristic} noise configuration, in which the $P$ and $A$ parameters are both reduced by a factor of $10$. 
}
\label{fig:degeneracy_scans2}
\end{figure}
Let us now explain a bit more in detail the second observation just mentioned.
At small primordial amplitude $\Omega_* <\Omega_{\mathrm{fg}}$ one can neglect the primordial contribution to
the covariance entering in the inner product definying the Fisher matrix:
\begin{equation}C_{OO'}\simeq \delta_{OO'}B_O \equiv \delta_{OO'}\left(M_{g,\,O}+N_O\right)\,,
\end{equation}
where we restored the channel indexes $O=\{A,E,T\}$.
For an overall amplitude parameter $\log\Omega_\ast$ (using $\partial/\partial\log\Omega_*\equiv \partial_\Omega$), we have for the primordial monopole part
$\partial_\Omega M_{p,\,A}\propto \mathcal{R}_A \,\Omega_\ast$ while he additional contributions, once the dipole is included in the analysis, enter only through $OT$ off-diagonals with $O=\{A,E\}$. Writing the dipole response schematically as  $D_{OT}\propto \beta\,\mathcal{R}_3\bar I$, we have
$\partial_\Omega D_{AT}\propto \beta\,\mathcal{R}_3 \Omega_\ast$.
Thus, in this low signal regime,
the monopole and dipole Fisher elements scales in the same way:
\begin{equation}
F^{\rm mono}_{\Omega\Omega}
\;\propto\;
\frac{(\partial_\Omega M_{p,\,A})^2}{B_A^2}
\;\propto\;\Omega_\ast^2\,,\qquad
F^{\rm dip}_{\Omega\Omega}
\;\propto\;
\frac{(\partial_\Omega D_{AT})^2}{B_A\,B_T}
\;\propto\;\Omega_\ast^2\,,
\end{equation}
and their ratio is approximately independent of $\Omega_\ast$. One can then write the ratio between the two marginalised variances\footnote{As seen in the previous section, assuming sufficiently strong priors, $\sigma_p\rightarrow 0$, the marginalised variances can be approximated simply by inverting the corresponding Fisher diagonal element.} at small signal amplitude as
\begin{equation}\label{ratio}
\frac{\sigma_{\log\Omega_*}^{\rm (with\ dip)}}{\sigma_{\log\Omega_*}^{\rm (no\ dip)}}\;\simeq\;\frac{1}{\sqrt{1+r}}\,,\qquad \mathrm{with}\,\,\,\,\, r \;\equiv\; \frac{F^{\rm dip}_{\Omega\Omega}}{F^{\rm mono}_{\Omega\Omega}}\,.
\end{equation}
For intuition, taking $B_A= B_E$, assuming that the $T$ channel is noise dominated, and drop the (often subdominant) $T$ contribution in the monopole denominator $B_T\simeq N_T$, the scaling of $r$ reduces schematically to
\begin{equation}\label{ratio2}
r \;\propto\;  \beta^2\frac{\mathcal{R}_3^2}{\mathcal{R}_A^2}\;\left(\frac{B_A}{B_T}\right)
\;\simeq\;
\text{const}\times \frac{M_{g,A}+N_A}{N_T}\,,
\end{equation}
where we remind that $M_{g,A}$ denotes the foreground contribution in the A (or E) channel.
If $M_{g,A}\gg N_A$, $r$ is controlled mainly by the foreground level. Thus the larger the foreground amplitude $\Omega_{\mathrm{fg}}$ the smallest the ratio in Eq. \eqref{ratio} (and so the wider the separation between the two curves``with" and ``without" dipole). For small foreground values,  $M_{g,A}\ll N_A$ , $r$ saturates to a value set mainly by the instrumental noise ratio $r \;\propto\; N_A/N_T$ that cannot beat the overall suppression outside the parenthesis in Eq. \eqref{ratio2}.

To assess how similar in shape the two components must be for the dipole to break the degeneracy, in Fig.~\ref{fig:degeneracy_scans2} we show the effect of varying the fiducial value of the primordial spectral index. As expected, the region of $n_t$ parameter space in which the inclusion of the dipole has a significant impact increases as the priors on the foreground spectral index become wider.

\subsection{Degeneracy with Noise}
As we explained, GWB measurements with LISA are complicated by the fact that noise is correlated among the three nested interferometers and, since the mission operates in a strong-signal regime, it is not possible to estimate the noise using data “far from the event” \cite{PhysRevD.109.042001}. This limited control over the noise can lead to degeneracies between the signal and imperfectly modelled noise contributions.
In this section, we show that the kinematic dipole can help break such degeneracies, particularly in situations where the noise template contains features that closely mimic the signal of interest. To illustrate this mechanism, we consider a toy model in which we introduce an \emph{additive, signal-shaped} noise component in the A/E/T auto-spectra, constructed to be (nearly) collinear with the monopole contribution of the GWB signal.
Writing the diagonal covariance as
\be
C_{OO}(f)=\,{\cal R}_O(f)\,\bar I(f)+N_O(f)\,,\ee
with
$
\bar I(f)\propto \,\Omega(f)/f^3
$
and $O\in\{A,E,T\}$, we modify the instrumental noise as 
\be
N_O(f)\to N_O(f)+\delta_{\text{noise}}\,S_O^{\rm fid}(f),
\ee
where the fixed template is chosen to match the fiducial signal shape:
\be
S_O^{\rm fid}(f)\equiv \,{\cal R}_O(f)\,\bar I_{\rm fid}(f)N_O(f_*),
\ee
where the constant value $N_O(f_*)$ has been added for reference.

\begin{figure}[h]
\centering
\includegraphics[width=0.49\textwidth]{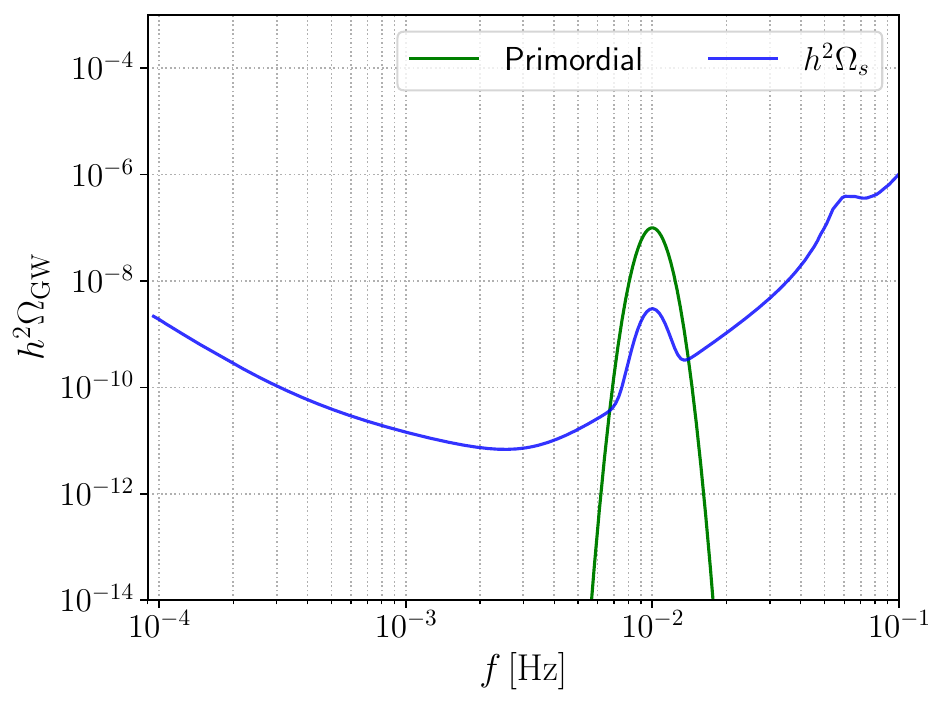}\quad
\includegraphics[width=0.47\textwidth]{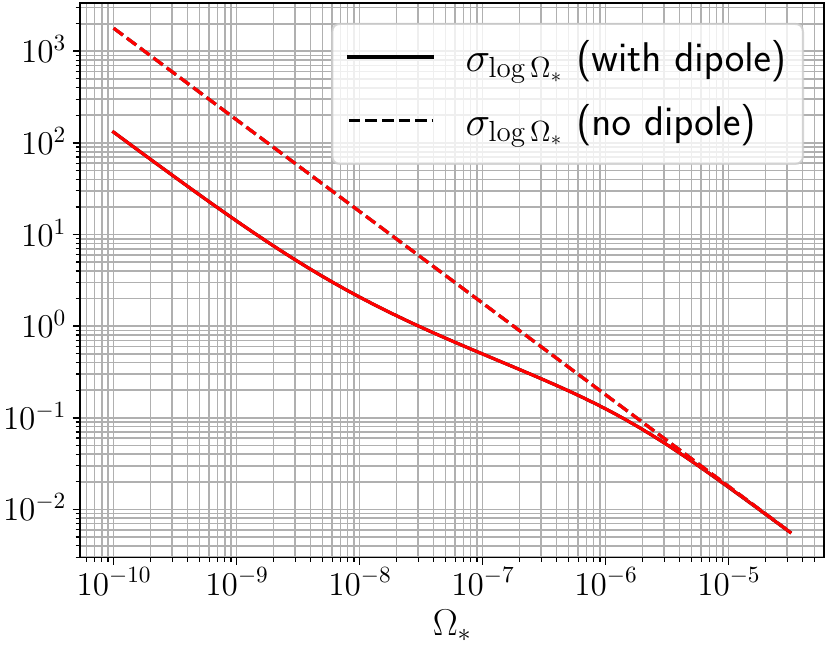}
\caption{Degenerate noise component. We consider a log-normal template for the signal,
$\Omega_{\mathrm{GW}}=\Omega_*\exp\!\left[-\tfrac{1}{2}\left(\ln(f/f_p)/\Delta\right)^2\right]$,
with parameters $(f_p,\Delta,\Omega_*)=(10\,\mathrm{mHz},0.1,10^{-7})$, and a noise deformation with the same functional form (see text), with fiducial values $(f_{p,\rm noise},\Delta_{\rm noise})=(10\,\mathrm{mHz},0.1)$. For the overall deformation parameter, we take $\delta_{\rm noise}=10^{-4}$ as fiducial and adopt a Gaussian prior with width $\sigma_p=10^{-2}$. Left: illustration of the benchmark signal. Right: scan over the signal amplitude. 
}
\label{fig:degeneracynoise}
\end{figure}

With the noise modification outlined above the derivatives entering in the Fisher matrix satisfy
$\partial C_{OO}/\partial \delta_{\text{noise}}=S_O^{\rm fid}(f)$,
which are proportional  to the derivatives with respect to the signal amplitude parameter, hence producing an (almost) exact degeneracy in the monopole-only Fisher analysis. In Fig.\,\ref{fig:degeneracynoise} we show the corner plot illustrating that the addition of a dipole significantly helps in constraining signal and noise parameters.

Note that this modification is somehow \emph{ad hoc}: not only the noise shape modification has to match the signal including the response function but we note that the same modification is applied to all three channels. Indeed, if the $T$ channel was not modified then for signals peaking at higher frequency in the LISA band (already at frequencies around $10\,\mathrm{mHz}$) the information coming from the signal in the $T$ channel would be sufficient to break the degeneracy between the overall primordial signal amplitude and the extra $\delta_{\text{noise}}$ parameter appearing in the modification of the noise in the A/E channels. 
In Fig.~\ref{fig:degeneracynoise}, we show the benchmark signal and illustrate how the variance evolves as a function of the signal amplitude, both with and without including the dipole in the analysis. For a fixed prior, the dipole ceases to improve the signal reconstruction once the signal exceeds a threshold amplitude, which increases as the prior is strengthened. In Fig.~\ref{fig:degeneracynoise}, we also present the corner plot for all the parameters entering the analysis. Note that the noise parameters, $A$ and $P$, are consistently well reconstructed, while the inclusion of the dipole improves the reconstruction of the other parameters. Notice that these results do not depend on the fiducial for the noise feature governed by $\delta_{\text{noise}}$ as it enters linearly in the analysis, hence the Fisher does not depend on its value.

\begin{figure}[h]
\centering
\includegraphics[width=0.77\textwidth]{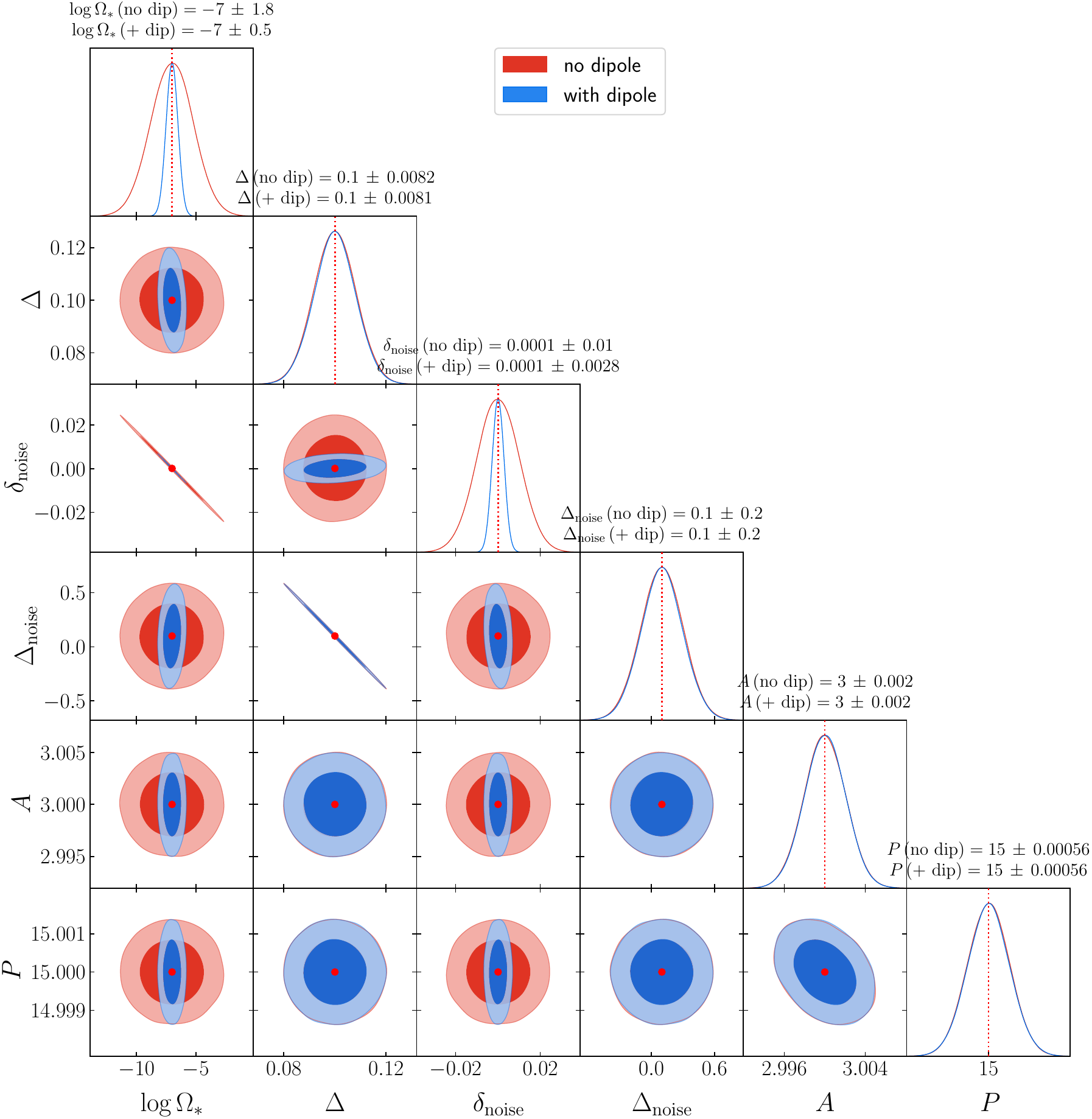}
\caption{Full corner plot for the benchmark model shown in the left panel of Fig.~\ref{fig:degeneracynoise}, where all parameters entering the analysis are displayed explicitly.}
\label{fig:degeneracynoise2}
\end{figure}

\section{Conclusion and Outlook}

In this work, we have developed an analytic framework to assess the capability of LISA to detect a kinematic dipole. This allows us to recover results previously obtained through numerical pipelines and to extend them to stochastic backgrounds with arbitrary spectral features. We show that the detectability of the kinematic dipole, and hence the reconstruction of our peculiar velocity, can be significantly more promising for backgrounds with richer spectral shapes. Finally, as a further novel application, we investigate the possible use of the kinematic dipole as a discriminator between foregrounds and instrumental noise.

We emphasize that our approach is simple, but not simplistic. Indeed, no approximations are introduced in the computation of the response function, apart from neglecting the breathing mode of the instrument, an assumption that is also commonly adopted in numerical studies. The advantage of our approach is that the underlying physics remains fully transparent, while forecasts can be derived very efficiently, without the need to run a complete numerical pipeline. Moreover, since our response functions are already integrated over the whole unit sphere, they could also be used to generate data streams that include kinematic anisotropies, thereby bypassing the need for a map-making algorithm. Such synthetic data could then be employed in analyses that go beyond the Fisher-forecast framework presented here.

As a non-trivial validation of our approach, we recover existing results in the literature for the variance of the velocity amplitude—previously obtained through simulations or numerical methods—\cite{Heisenberg:2024var, LISACosmologyWorkingGroup:2022kbp}.
For a scale-invariant spectrum, and assuming that the direction of motion is known and fixed to the value inferred from CMB measurements, we find that detectability can be achieved for $\Omega_* \gtrsim 5\times 10^{-8}$ with \emph{fiducial} LISA, and for $\Omega_* \gtrsim 5\times  10^{-10}$  for a \textit{futuristic} instrument with noise parameters reduced by an order of magnitude.
Our results for \emph{fiducial} LISA are in excellent agreement with the simulations of \cite{Heisenberg:2024var}. We also show analytically
that the inclusion of kinematic quadrupole anisotropies does not lead to an improvement in the constraints on the kinematic parameters.

A major challenge for space-based missions targeting a primordial background is the presence of strong astrophysical foregrounds. Furthermore, instrumental noise is expected to be less well characterized than in ground-based experiments, potentially leading to degeneracies between noise features and the signal. In this context, we have proposed the kinematic dipole as a robust degeneracy-breaking observable, since only the cosmological signal is modulated by the observer's motion.

We find that the dipole becomes particularly powerful in the presence of very large foregrounds, exceeding standard astrophysical expectations. Owing to the scale-invariant nature of the problem, however, our conclusions extend directly to future detectors with significantly improved sensitivity, even in scenarios where foregrounds remain large but astrophysically motivated. To the best of our knowledge, this is the first work in which the idea of using the kinematic dipole as a degeneracy breaker is explicitly proposed. We stress that our framework is intentionally simplified and should be regarded as a proof of concept. Several potentially relevant effects, such as intrinsic anisotropies, are neglected for the sake of illustration.

In our analysis, we mainly had in mind backgrounds of cosmological origin. In this context, intrinsic anisotropies are expected to be of the same order as those of the CMB, i.e., $(\sim 10^{-5})$, and are therefore strongly suppressed with respect to the kinematic modulation. Moreover, while the kinematic dipole depends on the spectral shape of the background, an intrinsic dipole does not. This difference could help distinguish the two contributions, particularly for backgrounds with non-trivial frequency dependence. For extragalactic backgrounds, intrinsic anisotropies may instead be comparable in amplitude to the kinematic dipole, depending on the underlying astrophysical model. In this case, one may exploit the fact that the direction of the kinematic dipole is accurately determined by CMB measurements (and is also consistent with the dipole inferred from the quasar distribution). Nevertheless, a detailed investigation of how to disentangle kinematic and intrinsic dipoles for astrophysical backgrounds is beyond the scope of the present work and deserves further study. We plan to address this issue in future work.

Furthermore, the limited usefulness of the quadrupole for velocity reconstruction is a result that emerges from our analysis. This finding is partly at odds with the general expectation (discussed below) that the quadrupole could provide a contribution of comparable size. We note that our conclusion that the quadrupole is not relevant for velocity reconstruction is consistent with the analysis of Ref.~\cite{Heisenberg:2024var}, which provides an indirect confirmation of our result. While Ref.~\cite{Heisenberg:2024var} includes both the dipole and the quadrupole, we reproduce their detectability estimates by including only the dipole, thereby explicitly verifying that the quadrupole contribution is negligible in this context.

Our study  provides a foundation for a systematic exploration of the science reach of post-LISA missions, and highlights the kinematic dipole as a promising tool for disentangling cosmological signals from foregrounds and instrumental systematics. A more detailed investigation of these aspects will be presented in future work.

\subsection*{Acknowledgments}
We are grateful to Jean-Baptiste Boyle, Chiara Caprini, Valerie Domcke,  Olaf Hartwig, Henri Inchauspe, Giorgio Mentasti, Mauro Pieroni and Subir Sarkar for interesting discussions during different stages of this project. The research of JF is supported by the
grant PID2022-136224NB-C22, funded by MCIN\allowbreak/\allowbreak AEI\allowbreak/10.13039\allowbreak/501100011033\allowbreak/\allowbreak FEDER,
UE, and by the grant\allowbreak/ 2021-SGR00872. 
GT is partially funded by the STFC grants ST/T000813/1 and ST/X000648/1. The work of GC and CP is supported by CNRS. The work
of GC has received financial support by the SNSF
Ambizione grant PZ00P2-193292,  \emph{Gravitational wave propagation in the clustered universe}. 

\appendix

\section{Response to a circularly polarized background}\label{SecPolarization}

Let us introduce the helicity basis components
\be
H_\pm(f,  \bn) = h_+(f,  \bn) \pm \ii h_\times(f,  \bn) 
\ee
If the GW background is partially circularly polarized, the statistics~\eqref{correlator} becomes
\be
\langle H_s (f,  \bn) H^*_{s'} (f',  \bn')\rangle\,=\,\delta_{s s'}\,\delta(f-f')\,\frac{\delta^{(2)} (\bn-\bn')}{4 \pi}\,I_s(f,\bn)\,,
\ee
with $I_+(f,\bn) \neq I_-(f,\bn)$, and the indices $s,s'$ take the values $\pm 1$. The correlation spectral density~\eqref{DefDXXi} is then replaced by 
\begin{equation}\label{DefDXXiC}
D_{XX'}(f;\tau_n)  = \sum_{s = \pm 1} \beta^i \, R^{\rm LISA}_{ij}(\tau_n) \, D^{j,s}_{XX'}(f)
\,(1-n^s_{I}(f))\,\bar{I}_s(f) \,.
\end{equation}
We find that $D^{i}_{XX'}(f) = 1/2 [D^{i,+}_{XX'}(f)+D^{i,-}_{XX'}(f) ]$ and
\begin{align}\label{RBetaC}
D^{i,{\cal C}}_{XX'}(f)\equiv\frac{1}{2} \left[D^{i,+}_{XX'}(f)-D^{i,-}_{XX'}(f) \right]= \ii {\cal R}_4(f) \ell^{o,\perp,i}_{X X'}
\end{align}
where for each pair $X,X'$, we defined the unit vectors 
\be
\ell_{AB}^{o,\perp,i} = \ell_{BC}^{o,\perp,i} = \ell_{CA}^{o,\perp,i} = -\ell_{BA}^{o,\perp,i} = -\ell_{CB}^{o,\perp,i} = -\ell_{AC}^{o,\perp,i} = e_z^i\,.
\ee
The parity odd response~\eqref{RBetaC} is sensitive only to the component of the peculiar velocity orthogonal to the LISA plane.\footnote{This is in agreement with \cite{Domcke:2019zls}, where, under the assumption that the kinematic dipole response functions depend only on the component orthogonal to the LISA plane, it was inferred from symmetry arguments at the level of the angular integrand defining the response functions that they are parity odd.} In the case of a non-chiral foreground, these contributions therefore sum to zero. By contrast, the contribution to the dipole response function computed in the main text, after summing over polarizations and assuming a non-chiral background, is parity even and depends on the component of the peculiar velocity lying in the LISA plane.
The low frequency approximation of the functional dependence is 
\begin{equation}
{\cal R}_4(f)=\frac{\sqrt{3}}{2} \left[-\frac{1}{5} + \frac{253}{3360}x^2 -\frac{187}{15129}x^4\right]+{\cal O}(x^6)\,.
\end{equation}
In the $A,E,T$ basis, using~\eqref{MagicPDP}, the response function~\eqref{RBetaC} takes the form 
\begin{equation}\label{RBetaAETC}
\left( D^{i,({\cal C})}_{O O'}(f)\right)=\ii \sqrt{3 }{\mathcal{R}}_4(f)\left(
\begin{array}{ccc}
0& e_z^i& 0\\
-e_z^i& 0& 0\\
0& 0& 0\\
\end{array}
\right)\,.
\end{equation}
As before, the full time dependence is restored via the rotation matrix involved in~\eqref{DefDXXiC}. In particular, this is done by including the time dependence of the normal to the LISA plane: ${\bm e}_z\rightarrow {\bm N}^{\rm LISA}(\tau_n) = R^{\rm LISA}(\tau_n)\cdot {\bm e}_z\,  $. Note also that after transforming to the AET basis, the parity-odd contributions to the dipole response function survive only in the $AE$ cross-correlation and cancel exactly in the other off-diagonal terms. This is precisely the opposite of what happens for the parity-even components, which cancel exactly in the $AE$ correlator (for an equilateral triangle) and survive instead in the $AT$ and $ET$ correlators; see Eq. \eqref{RBetaAET}.

\section{Description of LISA motion}\label{App:RLISA}

In Ecliptic coordinates at linear order in the eccentricity, these are the coordinates of each spacecraft~\cite{Rubbo:2003ap}
\begin{align}\label{ecliptic1}
x&=R \cos\alpha+\frac{1}{2} e R \left(\cos (2\alpha-\beta)-3 \cos\beta\right)\,,\\
y&=R \sin\alpha+\frac{1}{2} e R \left(\sin (2\alpha-\beta)-3 \sin\beta\right)\,,\\
z&=-\sqrt{3} e R \cos(\alpha-\beta)\,.\label{ecliptic3}
\end{align}
Here $\alpha=\Omega t+ \kappa$ is the
orbital phase of the guiding center and $\beta=2 \pi n/3 +\lambda$ where $n=0, 1, 2$ is the relative phase of the spacecraft within the constellation. The parameters $\kappa$ and $\lambda$ give the initial
ecliptic longitude and orientation of the constellation. The coordinates~\eqref{ecliptic1} can be rewritten in a geometrically more transparent form as
\be\label{xyzexplicit}
\left(\begin{array}{c}
x\\
y\\
z\\
\end{array}\right) = R_z(\alpha) \cdot \left(\begin{array}{c}
R\\
0\\
0\\
\end{array}\right) + R^{\rm LISA}(t) \cdot R_z(\beta) \cdot \left(\begin{array}{c}
-2 e R\\
0\\
0\\
\end{array}\right)
\ee
where 
\be
R^{\rm LISA}(t) \equiv R_z(\alpha) \cdot R_y (-\pi/3) \cdot R_z(-\alpha)\,.
\ee
The first contribution in~\eqref{xyzexplicit} is the position of the center of LISA constellation, whereas the second one is the rotation $R^{\rm LISA}(t)$ applied to the relative position of a satellite with respect to this center in the LISA frame.

\section{Fisher components}\label{FisherDet}

We provide here explicit expressions for the Fisher matrix (with LISA motion already accounted for) for three case study. First, we aim to simultaneously constrain parameters of a galactic and of an extragalactic components  (either astrophysical or cosmological).  Second, we want to  constrain parameters of a noise template and of a generic extragalactic signal.  In these two cases we keep dipole velocity fixed, assuming it to be the one measured by CMB experiments.

Latin lowercase letters denote derivatives with respect to parameters of an extragalactic background while greek letters are parameters of galactic background/noise. The integral over the entire range of frequency is understood in the rest of this section. 

\subsection{Constraining  cosmological and galactic parameters}

Let us assume we want to constrain parameters of an extragalactic background components. Then for the signal-signal part of the Fisher we get (with upper case indices running over the diagonal channels $A, E$ and $T$) 
\begin{align}\label{ssapp}
{F}_{pq}&=T \sum_O \int_{0}^{+\infty}\dd f\left(\frac{{\cal{R}}_O(f)}{M_O+N_O}\right)^2 \partial_p \bar{I}(f) \partial_q \bar{I}(f)\\
&+T \sum_{OO'} \int_{0}^{+\infty}\dd f\frac{1}{(M_O+N_O)(M_{O'}+N_{O'})}{\mathcal{T}}_{ijmn}\beta^i \beta^m D^j_{OO'}(f) D^n_{O'O}(f)\partial_p((1-n_I)\bar{I}) \partial_q((1-n_I)\bar{I}) \,,\nn
\end{align}
where
\be\label{Tau}
{\mathcal{T}}_{ijmn}=\langle R_{ij}^{\text{LISA}}(\tau)R_{mn}^{\text{LISA}}(\tau)\rangle_{\tau}\,.
\ee
Then we use the results of section \ref{LISAMoves} and we can simplify the second line of (\ref{ssapp})  using 
\begin{align}\nonumber
\sum_{OO'} &{\mathcal{T}}_{ijmn} \frac{D_{OO'}^{j} D_{O'O}^{n}}{(N_O+M_O) (N_{O'}+M_{O'})}=2\sum_{O} {\mathcal{T}}_{ijmn} \frac{D_{OT}^{j} D_{TO}^{n}}{(N_T+M_T) (N_O+M_O)}\\\nonumber
&= \frac{3{\mathcal{R}}_3^2(f)}{(N_T+M_T) (N_A+M_A)}\left[\delta_{im}-\langle N_{\text{LISA}}^i N_{\text{LISA}}^m\rangle_{\tau}\right]\\&=\frac{3{\mathcal{R}}_3^2(f)}{(N_T+M_T) (N_A+M_A)}\left[\frac{5}{8}\delta_{im}+\frac{1}{8} e_z^i e_z^m\right]\,,
\end{align}
where we used that $N_A=N_E$ and $M_A=M_E$ and going to the last equality we used (\ref{NLISAAv}). 
Then Eq.\,(\ref{ssapp}) simplifies to 
\begin{align}
{F}_{pq}=&T \sum_{O} \int_{
0}^{+\infty} \dd f\left(\frac{{\cal{R}}_O(f)}{M_O+N_O}\right)^2 \partial_p \bar{I}(f) \partial_q \bar{I}(f)\nn\\
&+T\cdot \frac{3}{8}\int_{0}^{+\infty} \dd f\frac{{\mathcal{R}}_3^2(f)}{(N_T+M_T) (N_A+M_A)}\left[5\beta^2+ \beta_z^2\right]\partial_p((1-n_I) \bar{I}(f))\partial_q((1-n_I) \bar{I}(f))\,.\label{sss}
\end{align}
We observe that the three velocity components in ecliptic coordinates do not enter in a symmetric way the expressions, with the  component normal to the ecliptic playing a special role. This was expected since LISA does not have the same resolution in all directions. 
For the off-diagonal block we get  
\be\label{sn2}
{F}_{p \alpha}=T\sum_O \int_{0}^{+\infty} \dd f\left(\frac{{\cal{R}}_O(f)}{(M+M_G)_O+N_O}\right)^2 \partial_p \bar{I}(f)  \partial_\alpha \bar{I}_g(f)\,,
\ee
where we have denoted with $M_G$ monopole contribution from the galactic foreground and with $\bar{I}_g$ the associated intensity. Finally, the galactic background block (equivalent of (\ref{nn})) writes 
\be
{F}_{\alpha \beta}=T \sum_O \int_{0}^{+\infty} \dd f\left(\frac{{\cal{R}}_O(f)}{(M+M_G)_O+N_O}\right)^2 \partial_{\alpha} \bar{I}_g(f) \partial_{\beta} \bar{I}_g(f)\,.
\ee

\subsection{Constraining noise and cosmological parameters}
Let us focus on  constraining extragalactic signal and noise parameters. The situation is very similar to what presented in the previous section. The signal-signal block is given by (\ref{sss}). 
The off-diagonal block
\be\label{sn}
{F}_{p \alpha}=T\sum_O \int_{0}^{+\infty} \dd f\frac{1}{(M_O+N_O)^2}\mathcal{R}_O(f) \partial_p \bar{I}(f) \partial_{\alpha}N_O(f)\,,
\ee
and finally the noise block
\be\label{nn}
{F}_{\alpha\beta}=T \sum_O \int_{0}^{+\infty} \dd f\frac{1}{(M_O+N_O)^2} \partial_\alpha N_O(f)\partial_{\beta}N_O(f)\,.
\ee

\section{Quadrupolar response function}\label{quadrupole}

In this Appendix we extend the discussion of Section~\ref{sec:dipolarresponse} to the kinematically-induced 
quadrupole. In fact,
at second order in the peculiar velocity, the Doppler boost in Eq.~\eqref{int_kin}
induces a quadrupolar modulation of the stochastic gravitational-wave background.
As a consequence, the correlation function in Eq.~\eqref{CgeneralXX} receives an additional contribution of the form
\begin{equation}\label{DefQXXi}
Q_{XX'}(f;\tau_n)  = \beta^i \beta^j \, R^{\rm LISA}_{ik}(\tau_n)R^{\rm LISA}_{jl}(\tau_n) \, Q^{kl}_{XX'}(f)\,(1-n^{Q}_I(f))\,\bar{I}(f) \, ,
\end{equation}
where $Q^{kl}_{XX'}(f)$ is the intrinsic quadrupolar response of the detector, and 
\begin{equation}
n^Q_I(f) = f \bar{I}'(f)- \frac{1}{2} f^2 \bar{I}''(f)\,.
\end{equation}
The response tensor is defined as
\begin{equation}
Q^{ij}_{XX'}(f) = \int \frac{\dd^2 \bn}{4\pi}
\sum_{\lambda=+,\times}
\left(n^i n^j - \frac{1}{3}\delta^{ij}\right)\,\Fresponse^{o,\lambda}_X(\bn,f)
\Fresponse^{o,\lambda \star}_{X'}(\bn,f)\, ,
\label{resp_quad}
\end{equation}
and is symmetric and traceless by construction.
In analogy with the dipole discussed in Section~\ref{sec:dipolarresponse}, the structure of Eq.~\eqref{resp_quad} is  constrained by symmetry. But the quadrupole transforms as a rank-2 tensor, and therefore spans a two-dimensional space of symmetric trace-free tensors in the LISA plane. As a result, the response cannot be described by a single function of frequency, but instead involves two independent scalar functions, multiplying two independent tensor structures.

A convenient description of kinematic quadrupole anisotropies is built in terms of  tensors
  aligned with the direction orthogonal to the LISA plane,
controlling anisotropies within the plane of the detector. 
For the autocorrelation channels, one finds
\begin{equation}
\label{exp_auAA}
Q^{ij}_{AA}(f) = {\cal Q}^{(0)}_1(f)\left(e_z^i e_z^j -\frac{1}{3}\delta^{ij}\right) +{\cal Q}_1^{(2)}(f) \left(\ell^{o,i}_{BC} \ell^{o,j}_{BC} - \tilde\ell^{o,i}_{BC}\tilde\ell^{o,j}_{BC}\right)\,,
\end{equation}
with analogous expressions for $BB$ and $CC$ obtained by cyclic permutations. Here $e_z^i$ denotes the unit vector orthogonal to the LISA plane, while $\tilde\ell_{BC}$ is a unit vector orthogonal to $\ell_{BC}$ within the plane. 
The second tensor multiplying ${\cal Q}_1^{(2)}$ in Eq.~\eqref{exp_auAA}  corresponds to a standard ``plus'' polarization pattern, which can be expressed entirely in terms of the LISA arm directions as
\begin{equation}
\left(\ell^{o,i}_{BC} \ell^{o,j}_{BC} - \tilde\ell^{o,i}_{BC}\tilde\ell^{o,j}_{BC}\right)
= \frac{4}{3}\left[
\frac12 \ell_{AB}^i \ell_{AB}^j
+
\frac12 \ell_{AC}^i \ell_{AC}^j
- \ell_{AB}^i \ell_{AC}^j
- \ell_{AC}^i \ell_{AB}^j
\right]\,,
\end{equation}
making explicit that the full response can be written solely in terms of the interferometer geometry.

For the cross-correlations, the structure is analogous:
\begin{equation}
Q^{ij}_{XY}(f) = {\cal Q}^{(0)}_2(f)\left(e_z^i e_z^j -\frac{1}{3}\delta^{ij}\right) +{\cal Q}_2^{(2)}(f) \left(\ell^{o,i}_{XY} \ell^{o,j}_{XY} - \tilde\ell^{o,i}_{XY}\tilde\ell^{o,j}_{XY}\right)\,.
\end{equation}
Hence,
the key difference with the dipole response  is that both diagonal and off-diagonal components are non-vanishing, reflecting the even-parity nature of the quadrupole.

The quadrupolar response is characterized by four functions of frequency -- two for the autocorrelation, two for the cross-correlations among channels (with $x \equiv f/f_\star$):
\begin{align}
{\cal Q}^{(0)}_1(f) &=\frac{3}{35}-\frac{13}{420}x^2+\frac{2921}{665\,280} x^4 -\frac{96\,403}{296\,524\,800} x^6+\frac{128\,633}{8\,717\,829\,120}x^8+{\cal O}(x^{10})\,,\\
{\cal Q}^{(0)}_2(f) &=-\frac{3}{70}+\frac{13}{840}x^2-\frac{2921}{1\,330\,560} x^4+ \frac{96\,403}{593\,049\,600} x^6 - \frac{563\,449}{87\,178\,291\,200}x^8+{\cal O}(x^{10})\,,\\
{\cal Q}^{(2)}_1(f) &=\frac{1}{504}x^2-\frac{661}{1\,330\,560}x^4 + \frac{49\,229}{1\,037\,836\,800}x^6 - \frac{271\,639}{108\,972\,864\,000}x ^8 +{\cal O}(x^{10})\,,\\
{\cal Q}^{(2)}_2(f) &=\frac{1}{504}x^2-\frac{1123}{1\,330\,560} x^4 + \frac{37\,823}{259\,459\,200}x^6- \frac{3\,073\,463}{217\,945\,728\,000}x^8+{\cal O}(x^{10})\,.
\end{align}

At low frequencies, the leading contribution arises from ${\cal Q}^{(0)}$, while ${\cal Q}^{(2)}$ is suppressed by additional powers of $x^2$. This behaviour reflects the reduced sensitivity of the detector to higher multipoles in the long-wavelength limit, consistently with the dipole scaling discussed in Section~\ref{sec:dipolarresponse}.

Despite its richer geometrical structure, the quadrupole gives only a limited
improvement for velocity reconstruction.  The reason is that two competing
effects are at work.  On the one hand, the quadrupole is a parity-even
anisotropy, to which the LISA response is comparatively more sensitive, and
its kinematic contribution involves higher derivatives of the monopole
spectrum.  This can enhance its frequency dependence, especially for spectra
with sharp features.  On the other hand, the quadrupole is generated only at
second order in the observer velocity.  Consequently, its contribution to the
Fisher information for the velocity is suppressed by powers of
$|{\bm \beta}|$.

For a fixed value of the velocity amplitude, and after averaging over the
orientation of ${\bm \beta}$, the quadrupolar contribution to the Fisher
matrix can be written as
\begin{equation}
F_{Q}^{\rm fixed}
=
|{\bm \beta}|^2 T
\int_0^\infty \dd f \, {\cal F}_Q(f)\,,
\end{equation}
where
\begin{align}
{\cal F}_Q(f)
&\equiv
\frac{16}{45}
\left[
2\left(
\frac{{\cal Q}^{(0)}_A (1-n^{Q}_I)\bar{I}(f)}
{{\cal R}_A \bar{I}(f) + N_A(f)}
\right)^2
+
\left(
\frac{{\cal Q}^{(0)}_T (1-n^{Q}_I)\bar{I}(f)}
{{\cal R}_T \bar{I}(f) + N_T(f)}
\right)^2
\right.
\nonumber\\
&\hspace{1.8cm}
\left.
+
3\left(
\frac{{\cal Q}^{(2)}_T (1-n^{Q}_I)\bar{I}(f)}
{{\cal R}_A \bar{I}(f) + N_A(f)}
\right)^2
+
6\,
\frac{
\left({\cal Q}^{(2)}_A (1-n^{Q}_I)\bar{I}(f)\right)^2
}{
\left({\cal R}_A \bar{I}(f) + N_A(f)\right)
\left({\cal R}_T \bar{I}(f) + N_T(f)\right)
}
\right] .
\end{align}
The quantity ${\cal F}_Q$ is the spectral Fisher density associated with the
quadrupolar part of the correlation function, with the explicit velocity
suppression factored out. The SNR associated to the detection of the velocity dipole via the quadrupole is related by
\be\label{SNRquadrupole}
\left.\text{SNR}_\beta^2 \right|_{\text{quad}} = |{\bm \beta}|^4 \int_0^\infty \dd f {\cal F}_Q(f)
\ee

In a purely noise-dominated regime, where the signal contribution in the
denominators can be neglected, one may find that
$|{\bm \beta}|^2{\cal F}_Q$ is only moderately smaller than the corresponding
dipolar contribution ${\cal F}_D$.  This observation alone could suggest that
the dipole and quadrupole enter velocity reconstruction on a comparable
footing; this expectation is also consistent with the relative behaviour of
the multipole sensitivity curves in Ref.~\cite{LISACosmologyWorkingGroup:2022kbp}.
However, this reasoning is misleading for the parameter range relevant to an
actual velocity measurement.

Indeed, the kinematic anisotropies are small. As seen on~\eqref{SNRdipole}, to constrain the velocity from
the dipole one needs, parametrically,
\begin{equation}\label{sqbD}
\left[
T\int_0^\infty \dd f\,{\cal F}_D(f)
\right]^{1/2}
\sim
\frac{1}{|{\bm \beta}|}
\sim
10^3 .
\end{equation}
From~\eqref{SNRquadrupole}, the requirement is even more demanding for the quadrupole whose amplitude is
of order $|{\bm \beta}|^2$, as we need
\begin{equation}\label{sqbQ}
\left[
T\int_0^\infty \dd f\,{\cal F}_Q(f)
\right]^{1/2}
\sim
\frac{1}{|{\bm \beta}|^2}
\sim
10^6 .
\end{equation}
Reaching this regime requires a sufficiently large monopole signal
$\bar I(f)$.  But once the monopole becomes large, it can no longer be
neglected in the covariance.  The denominators in the Fisher spectral
densities are then not controlled by the instrumental noises alone, but by
the combinations
\begin{equation}
{\cal R}_A\bar I(f)+N_A(f),
\qquad
{\cal R}_T\bar I(f)+N_T(f).
\end{equation}
Thus the same large signal that enhances the quadrupolar numerator also
increases the variance. This signal contribution to the covariance
compensates the apparent gain from the higher spectral derivatives and from
the favourable parity-even LISA response. Said differently, in the signal dominated regime both ${\cal F}_D(f)$ and ${\cal F}_Q(f)$ are of order unity and the expressions in brackets in~\eqref{sqbD} and~\eqref{sqbQ} are both of order $T f_{\rm max} \simeq 10^7$.

As a result, once the full signal-plus-noise covariance is retained, the
quadrupolar contribution remains subdominant in the velocity Fisher matrix,
schematically
\begin{equation}
F_{\beta}
\simeq
T\int_0^\infty \dd f\,{\cal F}_D(f)
+
|{\bm \beta}|^2 T\int_0^\infty \dd f\,{\cal F}_Q(f),
\qquad
|{\bm \beta}|^2{\cal F}_Q \ll {\cal F}_D
\end{equation}
throughout the regime in which the velocity can be constrained.  This
explains why, despite its richer tensor structure and its enhanced
sensitivity to parity-even multipoles, the quadrupole does not significantly
improve the velocity constraints in our setup.  Our analytical results are
in fact consistent with the simulation-based analysis of
Ref.~\cite{Heisenberg:2024var}, which also includes the quadrupole.

\bibliographystyle{JHEP}
\bibliography{LISAKINrefs}

\end{document}